\documentclass[fleqn,usenatbib]{mnras}

\usepackage{newtxtext,newtxmath}

\usepackage[T1]{fontenc}

\DeclareRobustCommand{\VAN}[3]{#2}
\let\VANthebibliography\thebibliography
\def\thebibliography{\DeclareRobustCommand{\VAN}[3]{##3}\VANthebibliography}

\usepackage{graphicx}	
\usepackage{amsmath}	
\usepackage{threeparttable}

\def\msun{M$_{\odot}$}
\def\rsun{R$_{\odot}$}

\def\degs{\ifmmode ^{\circ}\else$^{\circ}$\fi}
\def\amin{\ifmmode ^{\prime}\else$^{\prime}$\fi}
\def\asec{\ifmmode ^{\prime\prime}\else$^{\prime\prime}$\fi}
   
\def\degs{\ifmmode ^{\circ}\else$^{\circ}$\fi}
\def\amin{\ifmmode ^{\prime}\else$^{\prime}$\fi}
\def\asec{\ifmmode ^{\prime\prime}\else$^{\prime\prime}$\fi}
\def\farcs{\hbox{$.\!\!^{\prime\prime}$}}  
\def\h{$^{\rm h}$}
\def\m{$^{\rm m}$}

\newcommand{\fermi}{\textit{Fermi}}
\newcommand{\sw}{\textit{Swift}}
\newcommand{\ergs}{erg~s$^{-1}$}
\newcommand{\flux}{erg~s$^{-1}$~cm$^{-2}$}
\def\4fgl{4FGL J1838.2+3223}
\def\src{J1838}

\def\fermi{\textit{Fermi}}
\def\eros{\textit{eROSITA}}
\def\gaia{\textit{Gaia}}
\def\ps{Pan-STARRS}

\usepackage{ulem}

\title[A flaring `spider' candidate 4FGL J1838.2+3223]{Nature of 4FGL J1838.2+3223: a flaring `spider' pulsar candidate}

\author[D. A. Zyuzin et al.]{
D. A. Zyuzin$^{1}$\thanks{E-mail: da.zyuzin@gmail.com (DAZ)},
A. Yu. Kirichenko$^{2,1}$,
A. V. Karpova$^{1}$,
Yu. A. Shibanov$^{1}$,
S. V. Zharikov$^{2}$, 
M. R. Gilfanov$^{3,4}$, \newauthor
\ C. Perez T\'ortola$^{5}$
\\
$^{1}$Ioffe Institute, Politekhnicheskaya 26, St. Petersburg 194021, Russia\\
$^{2}$Instituto de Astronom\'ia, Universidad Nacional Aut\'onoma de M\'exico, Apdo. Postal 877, Ensenada, Baja California, M\'exico, 22800\\
$^{3}$Space Research Institute, Russian Academy of Sciences, Profsoyuznaya 84/32, Moscow 117997, Russia\\
$^{4}$Max-Planck-Institut f\"ur Astrophysik, Karl-Schwarzschild-Str. 1, D-85741 Garching, Germany\\
$^{5}$Instituto de Investigación en Ciencias Físicas y Matemáticas, USAC, Ciudad Universitaria, 01012, Zona 12, Guatemala \\
}

\date{Accepted XXX. Received YYY; in original form ZZZ}

\pubyear{2023}

\begin{document}
\label{firstpage}
\pagerange{\pageref{firstpage}--\pageref{lastpage}}
\maketitle

\begin{abstract}
An unidentified $\gamma$-ray source \4fgl\ has been  
proposed as a pulsar candidate. 
We present optical time-series multi-band photometry of 
its likely optical companion 
obtained with the 2.1-m telescope 
of Observatorio Astron\'omico Nacional San Pedro M\'artir, Mexico. 
The observations and the data from the Zwicky Transient Facility revealed the 
source brightness variability  
with a period of $\approx$4.02~h likely associated with the orbital motion of the binary system.
The folded light curves   have a single sine-like peak per period 
with an amplitude of about 
three magnitude accompanied by  fast sporadic flares  up to one magnitude level. 
We reproduce them 
modelling the companion heating by the pulsar. 
As a result, the companion side facing 
the pulsar is strongly heated up to 11300$\pm$400 K, while the temperature of its back side is only 2300$\pm$700 K. 
It has a  mass of 0.10$\pm$0.05~\msun\ and  underfills its Roche lobe with a filling factor of $0.60^{+0.10}_{-0.06}$. This implies that \4fgl\ likely belongs to the 
`spider' pulsar family. The estimated distance 
of $\approx$3.1 kpc is compatible with \gaia\ results.
We  detect a flare from the source in X-rays and ultraviolet   using  \sw\ archival data 
and another one in X-rays with the \eros\ all-sky survey.
Both flares have X-ray luminosity of $\sim$10$^{34}$~\ergs\ which is two orders of magnitude higher than the upper limit in quiescence obtained 
from \eros\ assuming spectral shape typical for spider pulsars. 
If the spider interpretation is correct, these flares are among the strongest flares observed from non-accreting spider pulsars.
\end{abstract}




\section{Introduction}

\begin{figure}
\begin{minipage}[h]{1.\linewidth}
\center{\includegraphics[width=1.\linewidth,clip]{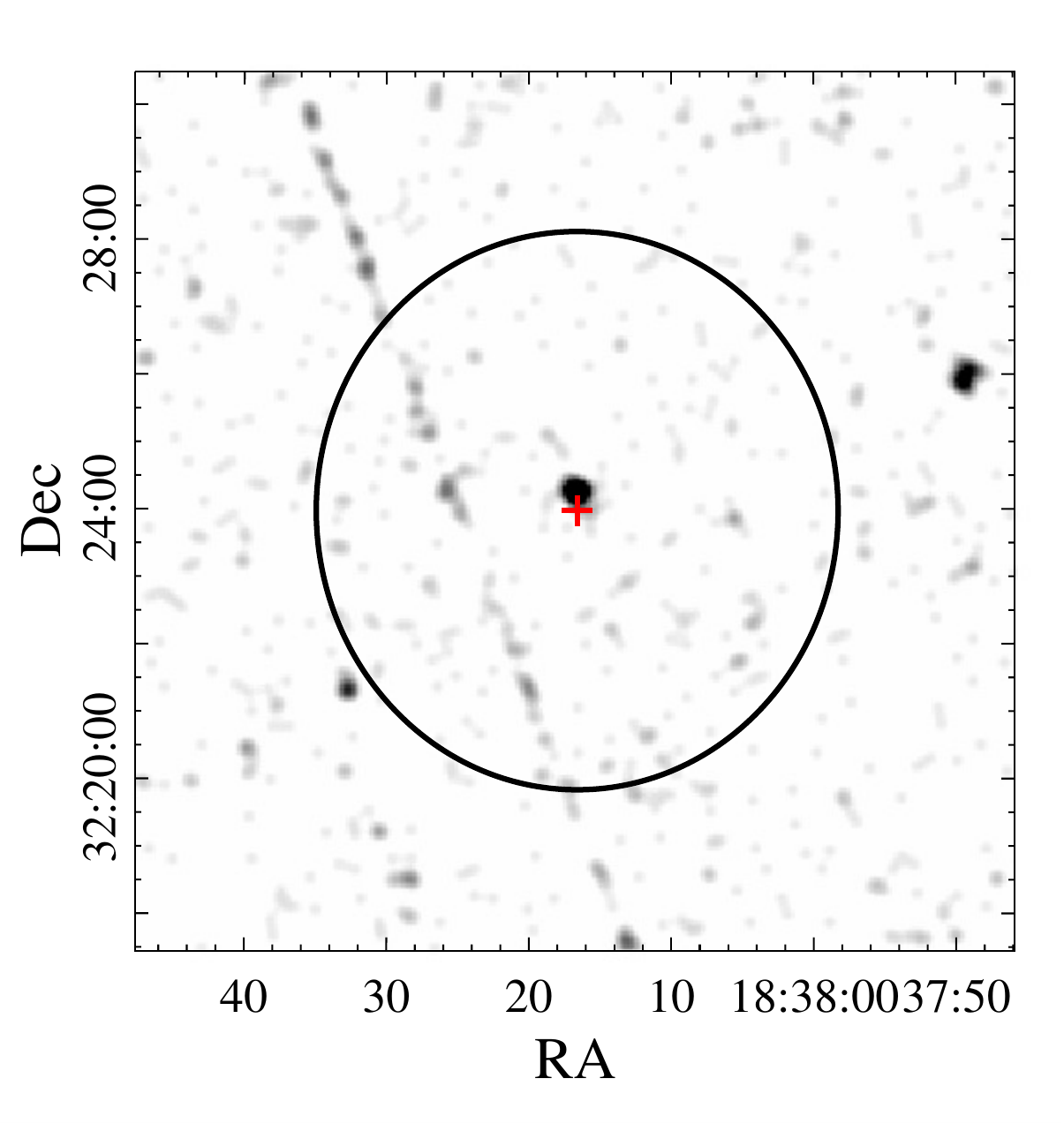}} 
\end{minipage}
\caption{13$\times$13 arcmin$^{2}$  
\sw/XRT image of  \src\ 
combined from different observations with the total exposure of 7.6 ks.  
The red cross and black ellipse show the $\gamma$-ray 
position of \src\ and its 
95 per cent uncertainty. The X-ray counterpart candidate proposed
by \citet{Kerby2021} is the  
nearest point object to the red cross. The  stripe crossing the field is an instrumental artifact.
}
\label{fig:ref-asked-for}
\end{figure}

\begin{figure*}
\setlength{\unitlength}{1mm}
\begin{picture}(120,78)(0,0)
\put(-25,0) {\includegraphics[width=8.5cm, clip=]{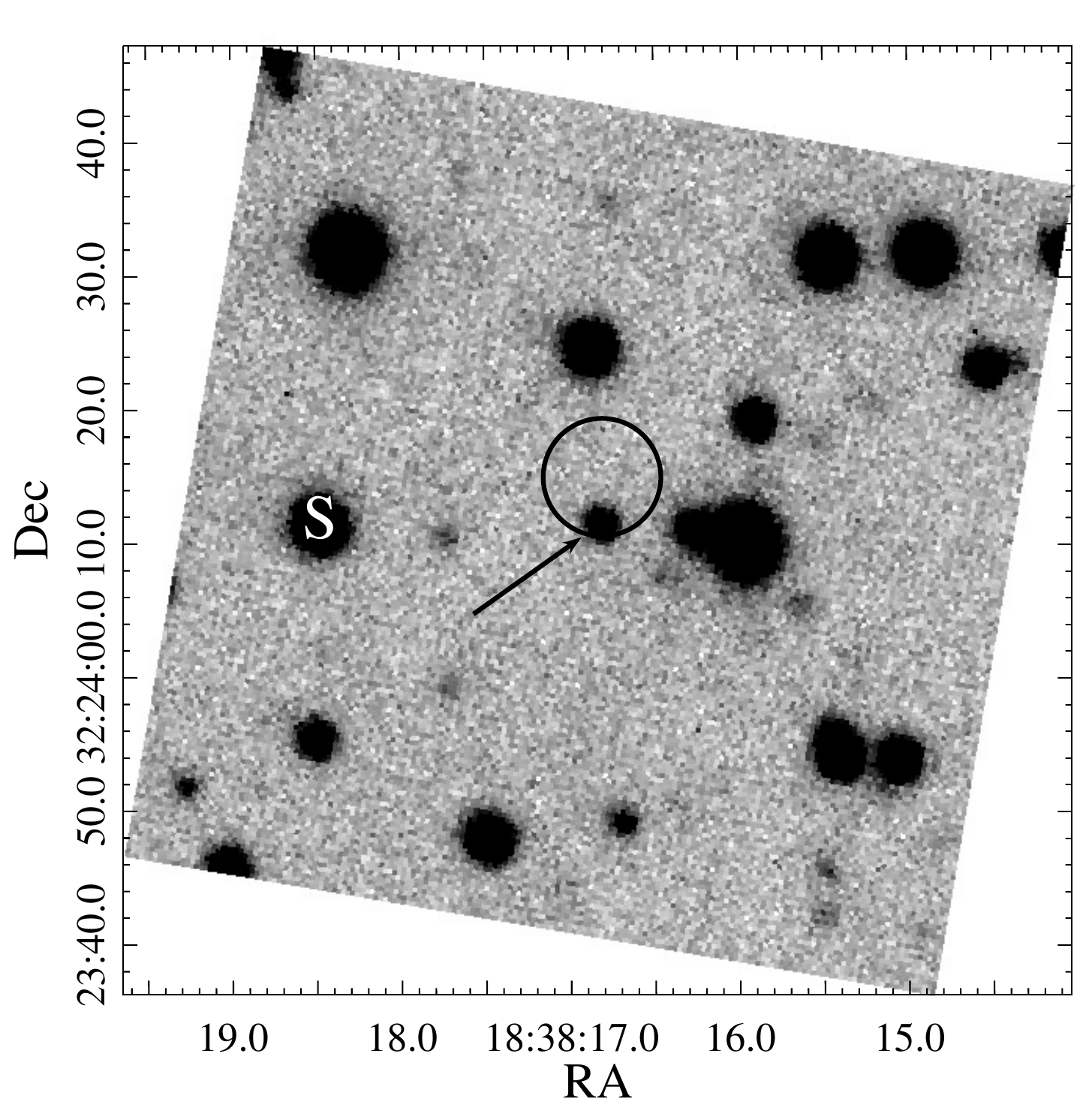}}
\put(60,0) {\includegraphics[width=8.5cm, clip=]{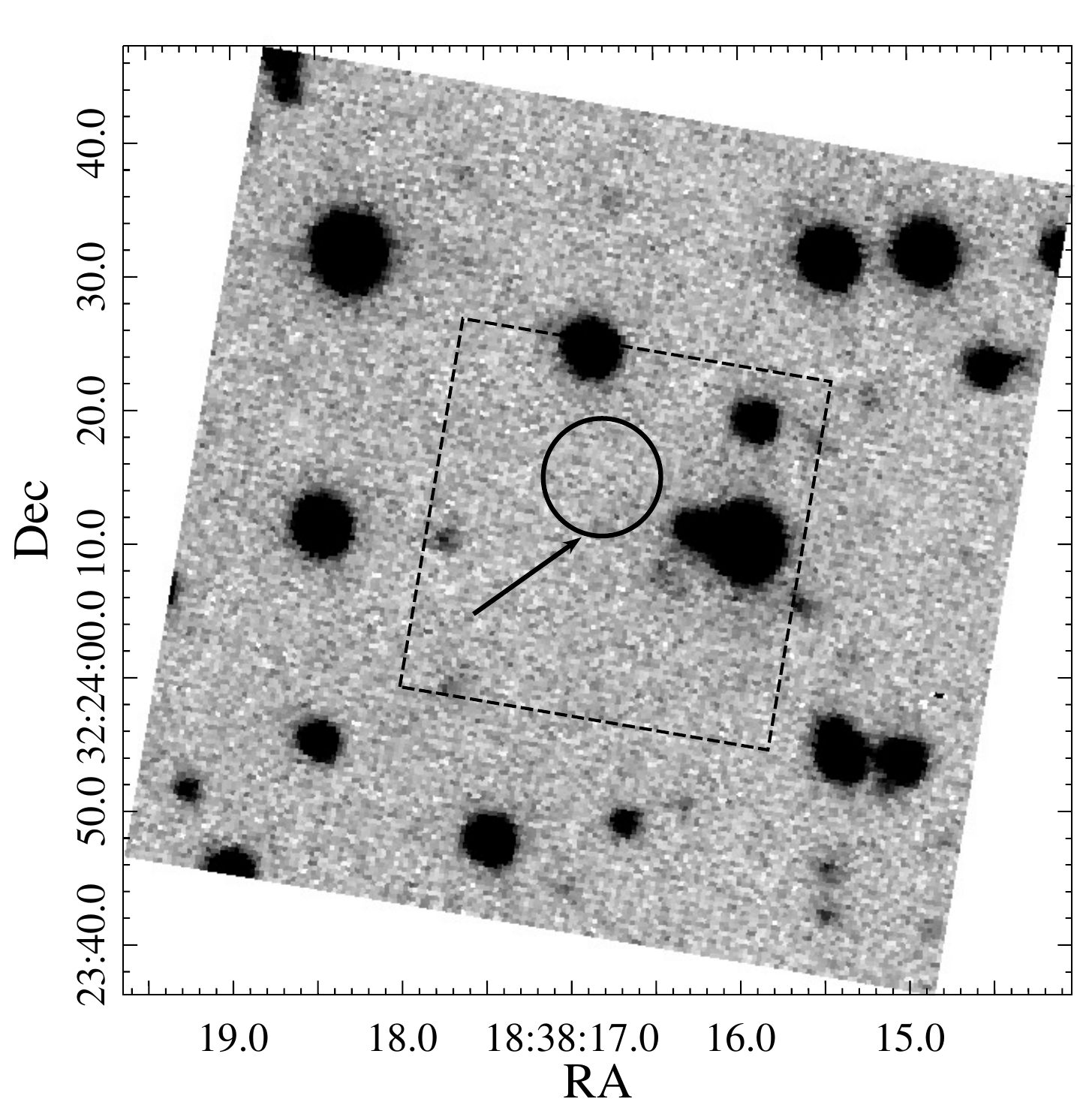}}
\put(117.5,59) {\frame {\includegraphics[width=2.25cm, clip=]{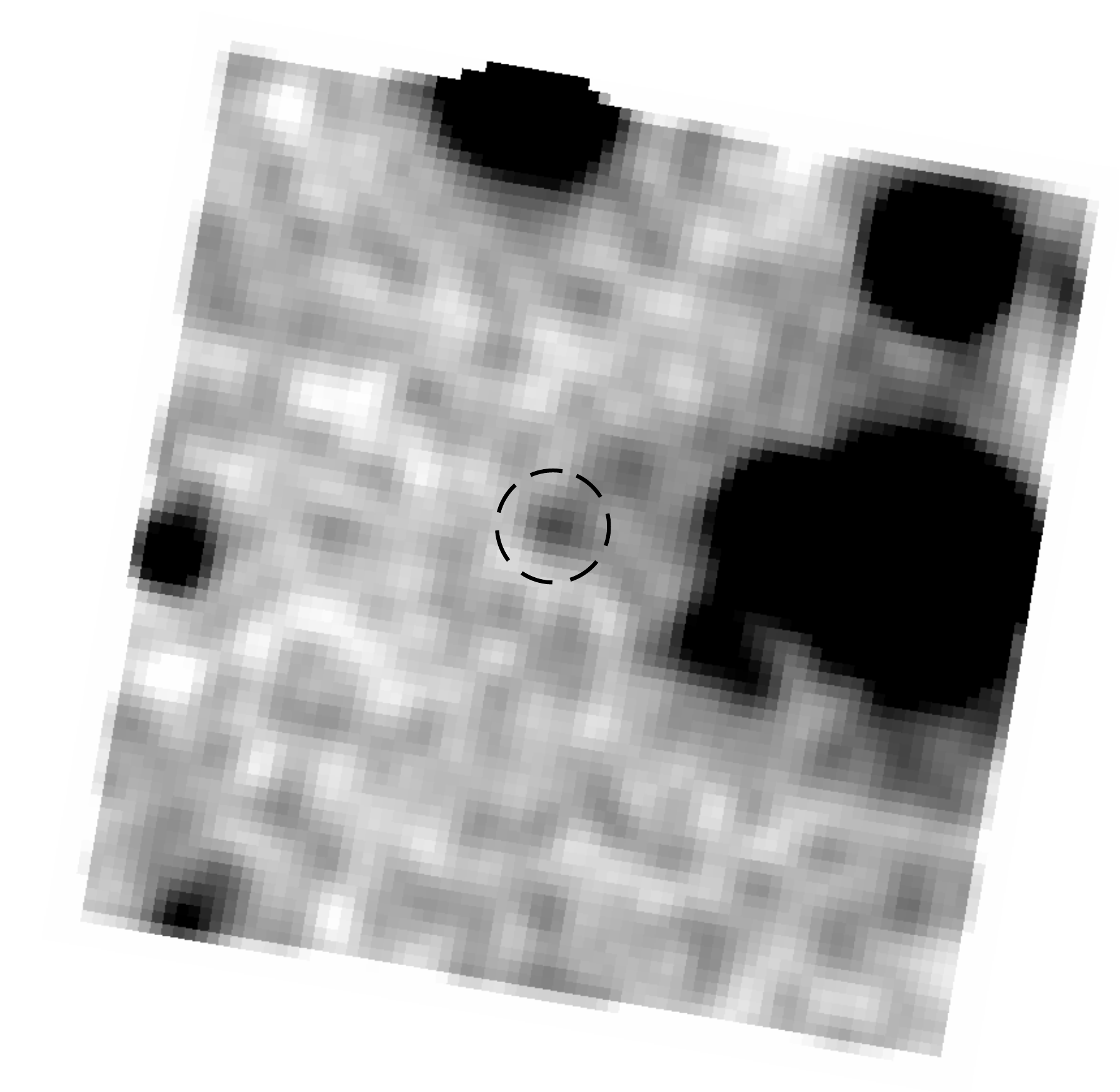}}}
\end{picture}
\caption{Individual $R$-band images of the \src\ 
field obtained with the OAN-SPM 
telescope during the maximum  
(left) and the minimum (right) brightness 
phases of the likely optical
counterpart marked by the arrow. 
The solid circle with a radius of 4.4 arcsec 
shows the \sw/XRT 
90 per cent position uncertainty of the \src\ X-ray counterpart candidate;  
`S' in the left panel denotes the star used as the secondary photometric standard in this work. 
The region enclosed in the dashed box in the right panel is shown in the inset 
with a different  intensity scale to better reveal 
the candidate (marked with the dashed circle) at the phase of its minimum  brightness.}
\label{fig:Rmaxmin}
\end{figure*}

To date, the \fermi\ Gamma-ray Space Telescope has detected 144
millisecond pulsars (MSPs) \citep{FermiPSRs2023}.
Among them, the subclass of so-called `spider' pulsars is of particular interest 
\citep{roberts2013}. It includes black widows (BWs) and redbacks (RBs) which are
compact binaries with orbital periods $P_b < 1$~d  where one side of 
the companion star is heated by the pulsar wind. 
BWs have very low-mass ($M_{\rm c}\lesssim 0.05$~\msun) semi-degenerate companions while RBs have non-degenerate and more massive ($M_{\rm c}$~$\approx$~0.1--1~\msun) companions.
Some RBs show transitions between radio-pulsar and accreting modes (e.g. \citealt{archibald2009,papitto2013,bassa2014,roy2015}) confirming the low-mass X-ray binaries as progenitors of spider systems.
However, details of formation and the evolutionary link between RBs and BWs are not clear and actively discussed   \citep{chen2013,benvenuto2014,ablimit2019,guo2022}. 
Studies of spider pulsars are also important for constraining  the equation of state of the superdense matter inside neutron stars (NSs) since the most massive of them have been found in binary systems (e.g. \citealt{romani2022}). 

Detection of pulsations from RBs and BWs is a somewhat challenging task since the pulsar emission is often eclipsed by material ablated from the companion star by the pulsar wind. 
However, even in the case of non-detection of pulsations, it is possible to identify such systems using optical and X-ray observations of unassociated $\gamma$-ray sources (e.g. \citealt{salvetti2017}). 
In particular, optical studies allow one to clarify the spectral type of the companion star, its mass and irradiation efficiency by the pulsar, the distance to the system, the binary inclination, and the pulsar mass. 
Only about four dozen confirmed spiders and about one dozen of candidates had been detected in the optical so far \citep{strader2019,miller2020,swihart2020,swihart2021,swihart2022,au2023,karpova2023,yap2023}. 
Parameters of many of them are poorly constrained due to the faintness of the objects. 
Thus, new identifications and studies of brighter companions are necessary for investigations of such systems. 

Recently, \citet{Kerby2021} built a neural network classifier trained on samples of known pulsars and blazars and used it to classify the unidentified \fermi\ sources.  Based on possible X-ray and optical counterparts  found with  the \sw\ X-ray Telescope (XRT) and UltraViolet Optical Telescope (UVOT), they identified 14 \fermi\ sources for which the probability 
of being a blazar was less than 1 per cent, thus making them likely pulsar candidates.
\4fgl\ (hereafter \src) is one of these sources. 
According to the 4FGL-DR3 catalogue, its $\gamma$-ray flux in the 0.1--100 GeV range is $F_\gamma=(2.7\pm0.6)\times 10^{-12}$~\flux\ \citep{4fgl-dr3}. 
The visual brightness of its  optical counterpart proposed by \citet{Kerby2021} is 21.3  mag. 
At the time of the \sw\ observation, the possible  X-ray counterpart was the brightest X-ray source in the \sw/XRT field of view (23.6$\times$23.6 arcmin$^2$) with the 0.3--10 keV  unabsorbed X-ray flux of $F_X \approx 7\times 10^{-13}$ \flux\ \citep{Kerby2021}. The proposed counterpart of \src\  is located 
in $\approx 20$ arcsec from the  \fermi\
position of J1838, and it is the only  X-ray source detected by \sw\ 
within the 95 per cent error ellipse of the $\gamma$-ray  
pulsar candidate \src\  (Fig.~\ref{fig:ref-asked-for}). 

To clarify the nature of the presumed X-ray and optical counterpart 
of \src, we performed multi-band optical observations and multi-epoch analysis 
of archival X-ray and optical data.
The data from catalogues are presented in Sec.~\ref{sec:cat}.
Optical observations and data reduction are described in Sec.~\ref{sec:data} 
and the light curves modelling -- in Sec.~\ref{sec:lc}. 
The description of the X-ray and ultraviolet (UV) data is presented 
in Sec.~\ref{sec:x-ray}. We discuss and summarise the results in Sec.~\ref{sec:discussion}.

\section{Data from catalogues}
\label{sec:cat}

The likely X-ray counterpart of \src\ suggested by 
\citet{Kerby2021} is nominated in their work as SwXF4 J183817.0+322416.  
Using the \sw/XRT 
data products generator\footnote{\url{https://www.swift.ac.uk/user_objects/}}  \citep{evans2009} we obtained its coordinates (see Table~\ref{tab:pars}) and the 90 per cent position uncertainty of 4.4 arcsec.
Within the  uncertainty circle we found  the   
presumed \src\ optical counterpart
in the Panoramic Survey Telescope and Rapid Response System 
(\ps, \citealt{ps2020}) with  ID 146882795700574433, which can also 
be cross-identified   
in \gaia\ \citep{gaia2016,gaia2021_edr3} and 
Zwicky Transient Facility (ZTF, \citealt{ztf2019}) archive data. 
Its parameters are presented in Table~\ref{tab:pars}.
In the \ps\ catalogue its mean AB magnitudes are
$g$ = 20.7(1), $r$ = 20.46(6), $i$ = 20.43(5) and $z$ = 20.5(1). 
These are apparently brighter than $m_V=21.3$ reported by \citet{Kerby2021}.  
However, based on individual \ps\ detections the source shows 
variability at a level of  1--2 stellar magnitude. 
The ZTF data confirm the variability.

The `geometric' distance $D_{\rm geom}$  obtained with \gaia\ lies 
in the range 1--3.5~kpc, while the 
`photogeometric' distance $D_{\rm pgeom}$ is 8.7--12.4 kpc.
The proper motion $\mu=11.1 \pm 1.4$ mas~yr$^{-1}$ 
transforms to the transverse velocity of about 50--200 km~s$^{-1}$ for $D_{\rm geom}$ and 500--700 km~s$^{-1}$ for $D_{\rm pgeom}$.
The latter is much larger than expected from the velocity
distribution for binary systems with pulsars \citep{hobbs2005}. 
This suggests
that the photogeometric distance
is unrealistic 
and we excluded it from consideration. 
For $D_{\rm geom}$, the \src\ $\gamma$-ray luminosity is $L_\gamma$~=~(0.3--4)$\times$10$^{33}$~\ergs\ 
and the X-ray luminosity obtained from the flux derived by \citet{Kerby2021} is $L_X \approx 10^{32}$--$10^{33}$ \ergs.
Such parameters are typical for the spider pulsar family \citep{strader2019,swihart2022}.

\begin{table}
\renewcommand{\arraystretch}{1.2}
\caption{Parameters of the \src\ and its counterpart candidate.}
\label{tab:pars}
\begin{center}
\begin{tabular}{lc}
\hline
\multicolumn{2}{c}{Parameters from the literature}                            \\  
R.A. in the optical $\alpha_{\rm opt}$                           & 18\h38\m16\fs81807(12)    \\
Dec. in the optical $\delta_{\rm opt}$                           & +32\degs24\amin11\farcs4148(11)\\
Galactic longitude $l$, deg                                      & 61.283                         \\
Galactic latitude $b$, deg                                       & 16.706                         \\
P.m. in R.A. direction $\mu_{\alpha}$cos$\delta$, mas yr$^{-1}$  & $-$4.9(2.0) \\
P.m. in Dec. direction $\mu_\delta$, mas yr$^{-1}$               & $-$9.9(1.2) \\
Distance $D_{\rm geom}$, kpc                                     & 1.0--3.5 \\
$\gamma$-ray flux $F_\gamma$, \flux                              & $2.7(6)\times 10^{-12}$ \\
\hline
\multicolumn{2}{c}{Parameters derived in this paper}                            \\
R.A. in X-rays $\alpha_X$                                        & 18\h38\m16\fs71    \\
Dec. in X-rays $\delta_X$                                        & +32\degs24\amin1\farcs8\\
Orbital period $P_b$, h                                          & 4.02488(15) \\
X-ray flux (flare) $F_X^{\rm flare}$, \flux\                     & $\sim 10^{-11}$ \\
X-ray flux (quiescence) $F_X$, \flux\                            & $<9\times 10^{-14}$  \\ 
X-ray luminosity (flare) $L_X^{\rm flare}$,                      & $\sim 10^{34}$ \\
$(D/3.1\ {\rm kpc})^2$ \ergs\                                    &            \\
X-ray luminosity (quiescence) $L_X$,                             & $<10^{32}$ \\ 
$(D/3.1\ {\rm kpc})^2$ \ergs\                                    & \\
$\gamma$-ray luminosity $L_\gamma$,                              & $3.1(7)\times$10$^{33}$ \\
$(D/3.1\ {\rm kpc})^2$  \ergs\                                   & \\
\hline
\end{tabular}
\end{center}
\begin{tablenotes}
\item Hereafter numbers in parentheses denote 1$\sigma$ uncertainties relating to the last 
significant digit. 
\item The optical coordinates and the proper motion (p.m.) are obtained from the \gaia\ catalogue. 
\item $F_X$ and $F_X^{\rm flare}$ are the unabsorbed fluxes in the 0.5--10 keV range and $F_\gamma$ is the flux in the 0.1--100 GeV range. 
\end{tablenotes}
\end{table}


\section{Optical observations and data reduction}
\label{sec:data}

\begin{table}
\caption{Log of the \src\ observations with the OAN-SPM 2.1-m telescope.} 
\label{log}
\begin{tabular}{ccccc}\hline
Date       & Filter & Number    & Airmass    & Seeing,  \\ 
YYYY/MM/DD           &        & of exposures &            & arcsec  \\
\hline
2022/07/21 & $B$    & 3         & 1.0$-$1.5   &  1.2--1.4       \\
           & $V$    & 18        & 1.0$-$1.3  &           \\
           & $R$    & 8         & 1.0$-$1.5   &          \\
           & $I$    & 4         & 1.0$-$1.4  &           \\
\hline
2022/07/22 & $B$    & 5         & 1.0$-$1.5  &   1.5--1.7         \\
 & $V$    & 6         &  1.0$-$1.3          &          \\
 & $R$    & 30        &  1.0$-$1.5          &          \\
 & $I$    & 6         & 1.0$-$1.3  &           \\
\hline
2022/07/24 & $B$    & 5         & 1.0$-$1.1  &  1.3--1.8         \\
 & $V$    & 5         & 1.0$-$1.2 &       \\
 & $R$    & 30        & 1.0$-$1.2 &            \\
 & $I$    & 5         & 1.1$-$1.2 &           \\
\hline
\end{tabular}
\end{table}

The time-series photometric observations of  \src\ were performed using the $BVRI$ Johnson-Cousins bands
with the `Rueda Italiana' instrument\footnote{\url{https://www.astrossp.unam.mx/en/users/instruments/ccd-imaging/filter-wheel-italiana}} 
attached to the 2.1-m telescope at the Observatorio Astron\'omico Nacional San Pedro M\'artir (OAN-SPM), Mexico in July 2022. 
The field of view of the detector is 6$\times$6 arcmin$^2$ 
with an image scale of $0.34$ arcsec in the 2$\times$2 CCD pixel binning mode. 
The log of observations is given in Table~\ref{log}. 
Individual exposures varied from 350 to 700~s.  

We carried out standard data processing, including bias subtraction 
and flat-fielding utilising 
the {\sc iraf} package. 
In addition, fringe correction was performed for all $I$-band images.
Astrometric solution was calculated 
using
a set of 9 stars from the \textit{Gaia} eDR3 Catalogue 
\citep{gaia2016,gaia2021_edr3,gaia-edr3-astrometry}  
with positional errors of $\lesssim$0.13 mas. 
Formal $rms$ uncertainties of the resulting astrometric fit were $\Delta\alpha$~$\lesssim0.04$~arcsec 
and $\Delta\delta$~$\lesssim0.05$~arcsec. 
The photometric calibration was performed using the Landolt standards SA 109-949, 954, 956  \citep{1992AJ....104..340L} observed on 2022 July 22. 
All the data from the other nights were re-calibrated using a field non-variable star (\ps\ ID 146882795769764297, $r=17.67(4)$ mag)  
close to the target as a secondary standard. It is marked by `S' in Fig.~\ref{fig:Rmaxmin}, left. 
According to \ps\, its brightness variability is $\Delta m \lesssim 0.1$ mag. 

\begin{figure}
\begin{minipage}[h]{1.\linewidth}
\center{\includegraphics[width=1.\linewidth,clip]{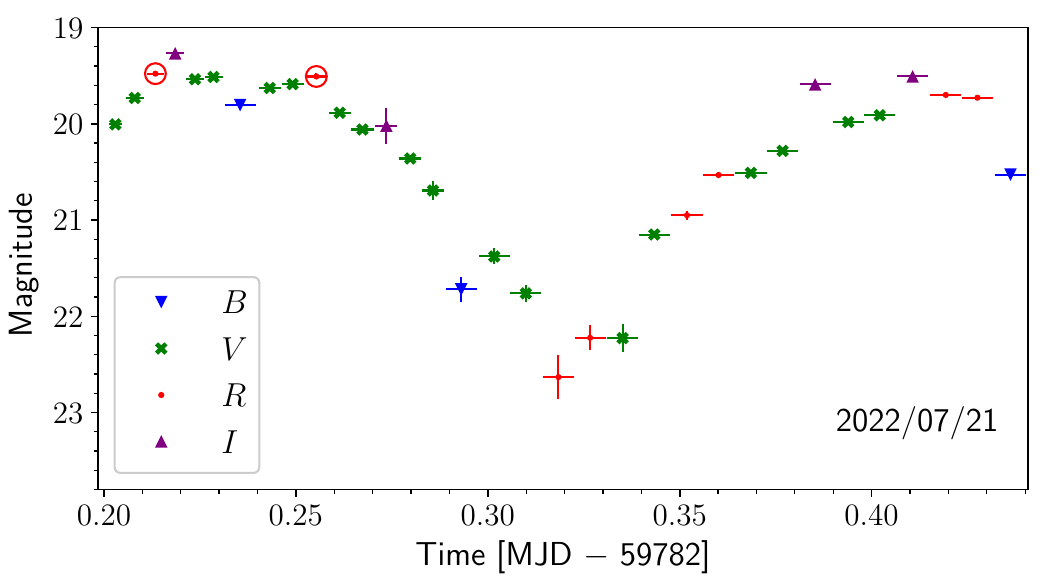}}
\center{\includegraphics[width=1.\linewidth,clip]{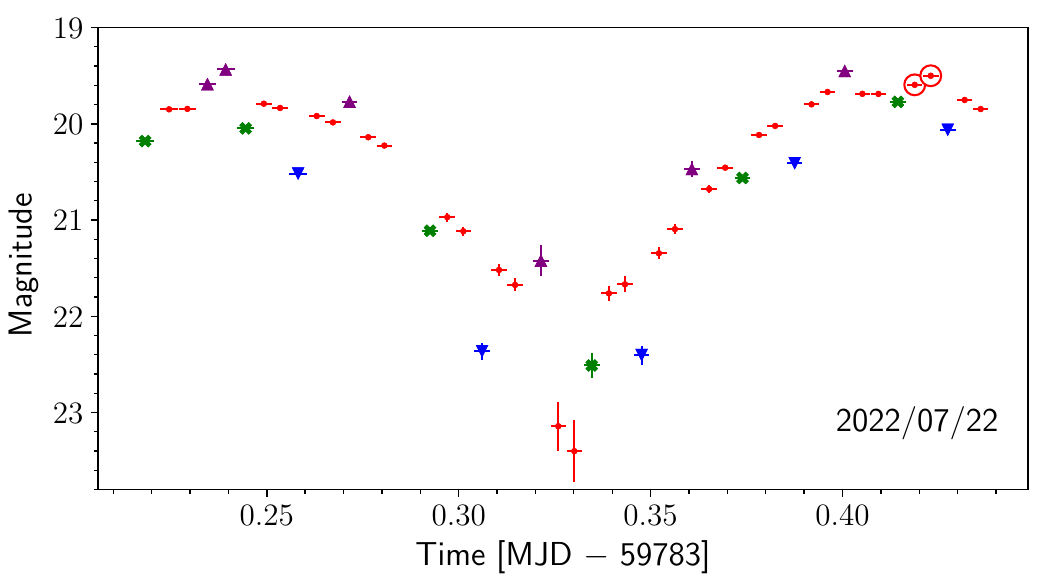}}
\center{\includegraphics[width=1.\linewidth,clip]{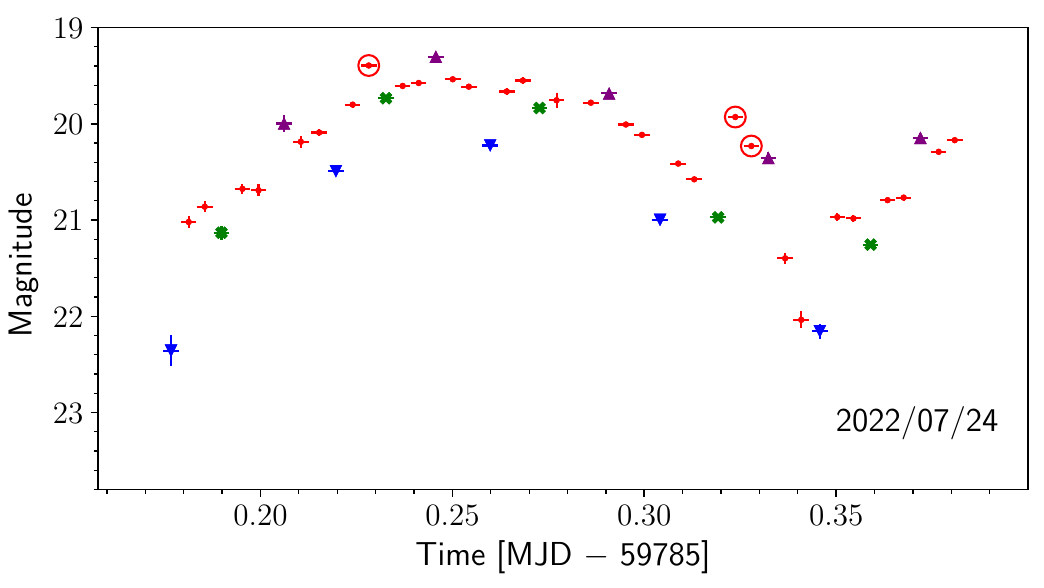}}
\end{minipage}
\caption{Time-series photometry of the \src\ putative optical counterpart 
obtained with the OAN-SPM 2.1-m telescope at different dates shown in the panels.
Points excluded from the timing analysis are marked with circles (see text).
}
\label{fig:lc-unfolded}
\end{figure}


\section{Images, light curves and orbital period}
\label{sec:lc}

Representative images  of the \src\ field obtained with  the OAN-SPM telescope in the $R$ band 
are shown in Fig.~\ref{fig:Rmaxmin}. They demonstrate that the \src\ likely optical 
counterpart is a highly variable source. 
It is firmly detected  in most exposures  and appears close to
the detection limit of $\approx$23.8 in some others. 
There is another faint object detected inside the X-ray position uncertainty and located about 2.5 arcsec north-west of the counterpart candidate 
(see the inset in  Fig.\ref{fig:Rmaxmin}, right). 
The source is also visible close to the detection 
limit in the Pan-STARRS images.  We did not find any variability 
associated with this background object in our data. 

The OAN-SPM light curves of the candidate in four bands    are presented in  Fig.~\ref{fig:lc-unfolded}. 
A smooth sine-like brightness variation with an amplitude of several magnitudes 
is accompanied 
by a short flaring activity on time-scales of $\sim$10 min with the brightness increase by 
about 1 mag. 
This short-term variability is especially prominent in the data obtained on 2022 July 24. 
The data  suggest a brightness periodicity of around 4~h. 
To find the period, we performed a Lomb-Scargle periodogram analysis \citep{lomb1976,scargle1982} 
applying the period range 1--24~h.
Excluding the most obvious flares in the $R$-band marked by the circles 
in Fig.~\ref{fig:lc-unfolded}, we obtained a period of 4.0246~h.

\begin{figure}
\begin{minipage}[h]{1.\linewidth}
\center{\includegraphics[width=1.\linewidth,clip]{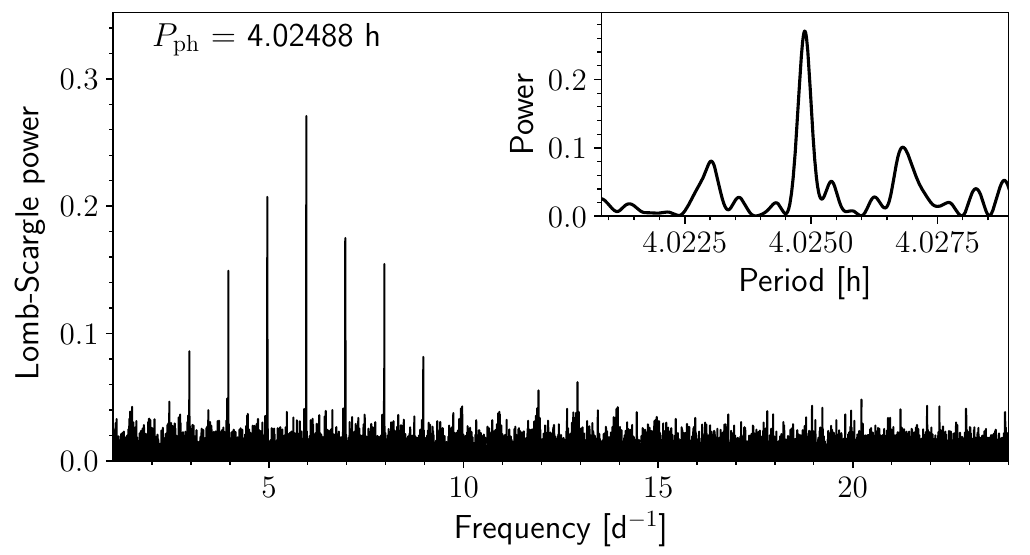}}
\end{minipage}
\caption{Lomb-Scargle periodogram for the ZTF data in the $r$ band calculated excluding points with $r<19.2$ mag.
The best period $P_{\rm ph}$ corresponding to the highest peak, enlarged in the inset, is also indicated.
}
\label{fig:period-ztf}
\end{figure}

To justify the period independently, we investigated the data from the ZTF DR16 catalogue
which covers about 4.7 yr (MJD 58204--59905) and contains several hundreds 
of photometric measurements of the source in the $r$-band.
The highest peak in the periodogram corresponds to the period  $P_{\rm ph}=4.02489$ h, 
which is very close to the value obtained from the OAN-SPM data.
However, the power spectrum is noisy, and folding the light curve 
with the obtained period we found indications of a heavy flaring activity, 
as has been revealed in the OAN-SPM data.
Thus, we excluded ZTF measurements with $r<19.2$ which are likely
related to the flares and recalculated the periodogram. 
The latter is shown in Fig.~\ref{fig:period-ztf}. 
The  highest peak in the power spectrum at $P_{\rm ph}=4.02488$ h is only slightly shifted from the  previous estimates and is quite pronounced. 
Since the $R$-band curve of the OAN-SPM contains only about 60 points, as opposed to the ZTF $r$-band, and covers a much shorter time range, 
we adopt  $P_{\rm ph}=4.02488(15)$~h\footnote{The uncertainty
is calculated as the half width at half maximum of the highest peak in the periodogram.}
obtained from the ZTF data as the true period of the \src\ candidate counterpart.

\begin{figure}
\begin{minipage}[h]{1.\linewidth}
\center{\includegraphics[width=1.\linewidth,clip]{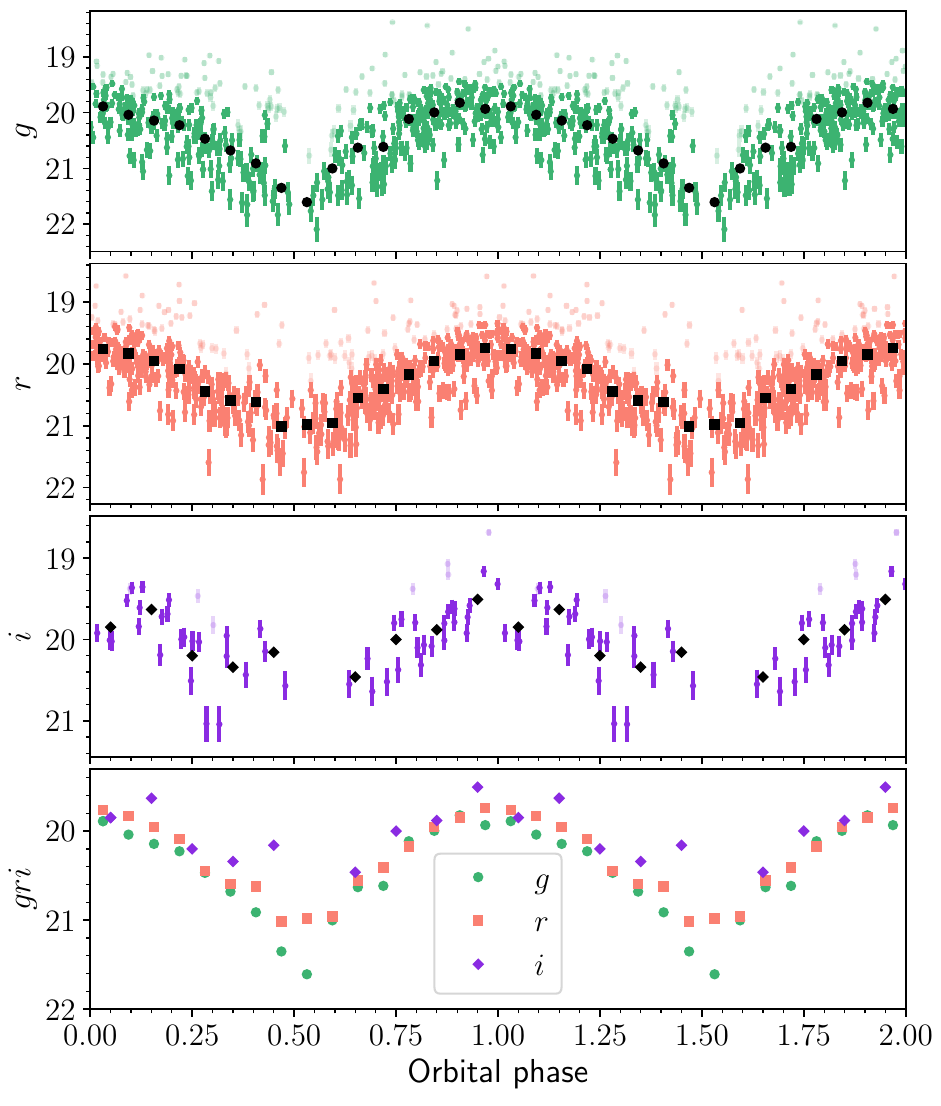}}
\end{minipage}
\caption{ZTF optical light curves of the \src\ presumed companion  in the $g$, $r$ and $i$ bands (panels 1--3) folded with a period of 4.02488 h.
The data points associated with flaring and excluded from analysis are
marked by lighter colours.
The mean binned
light curves are marked by black symbols in panels 1--3 and are shown together in panel 4. }
\label{fig:lc-ztf}
\end{figure}

The ZTF  light curves in the $g$, $r$ and $i$ bands folded with this period are shown in Fig.~\ref{fig:lc-ztf}. 
They demonstrate a single broad peak per period. 
We calculated the mean binned ZTF light curves and removed all data points with magnitudes 0.3
mag lower than the mean value in each phase bin since they are likely caused
by the flaring activity. 
After that, we calculated the
mean binned light curves again - they are compiled together in the bottom panel of Fig.~\ref{fig:lc-ztf}.
The minima of the ZTF light curves at the orbital phase 0.5 
have no detection points. Averaging over larger 
phase bins artificially smooths  and decreases the depth of the minimum  (see the bottom panel of Fig.~\ref{fig:lc-ztf}).

\begin{figure}
\begin{minipage}[h]{1.\linewidth}
\center{\includegraphics[width=1.\linewidth,trim={0 0 0 0.38cm},clip]{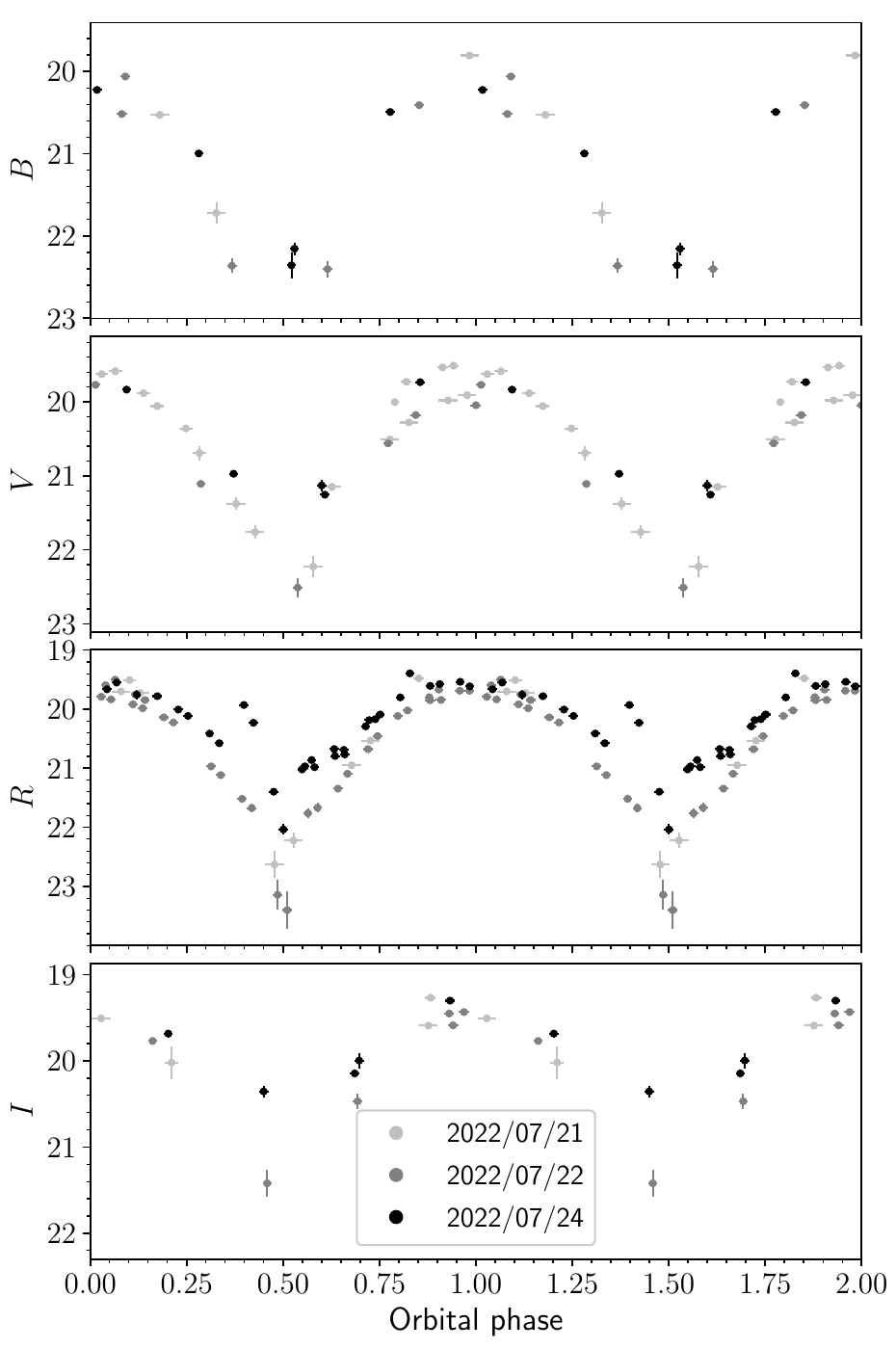}}
\end{minipage}
\caption{Light curves of the \src\ likely counterpart obtained with  the 2.1-meter telescope folded with a period of 4.02488 h. The data obtained during different nights are marked by the different grey colours as indicated in the legend.}
\label{fig:lc-BVRI}
\end{figure}

The light curves obtained with the OAN-SPM  
are demonstrated in Fig.~\ref{fig:lc-BVRI}. 
In this case, deeper observations allowed us to measure the light curves 
minima more accurately resulting in a total brightness variation 
of about 3 mag (from $\approx 20$ to $\approx 23$ mag) in the $R$ band. 

The period value and shapes of the periodic light curves 
are typical for a spider pulsar companion owing to its  orbital motion around a MSP \citep[e.g.][]{matasanchez2023,kandel2020,linares2018}. 
We thus can assume that $P_{\rm ph}$ is an orbital period of \src. 
We  note, that on 2022 July 24 the source was apparently 
brighter throughout the whole observation
as compared to the two previous nights. This is particularly 
clearly seen in the $R$-band light curve 
in Fig.~\ref{fig:lc-BVRI} at the phase ranges between about 0.2 and 0.8.  
It may reflect a global change of the presumed pulsar 
companion and/or the binary system  stage.


\subsection{Light-curve modelling}
\label{sec:lc-mod}

To estimate the  parameters of the presumed binary system, we performed the light curve modelling
using the technique described in \citet{zharikov2013,zharikov2019}.
We modelled only  the OAN-SPM data because they are deeper than the ZTF ones and consequently cover the whole  period per each observing night.
We excluded the data obtained on 2022 July 24 when \src\ 
was apparently in a different brightness stage.

\begin{figure*}
\begin{minipage}[h]{1.\linewidth}
\includegraphics[width=0.5\linewidth,clip]{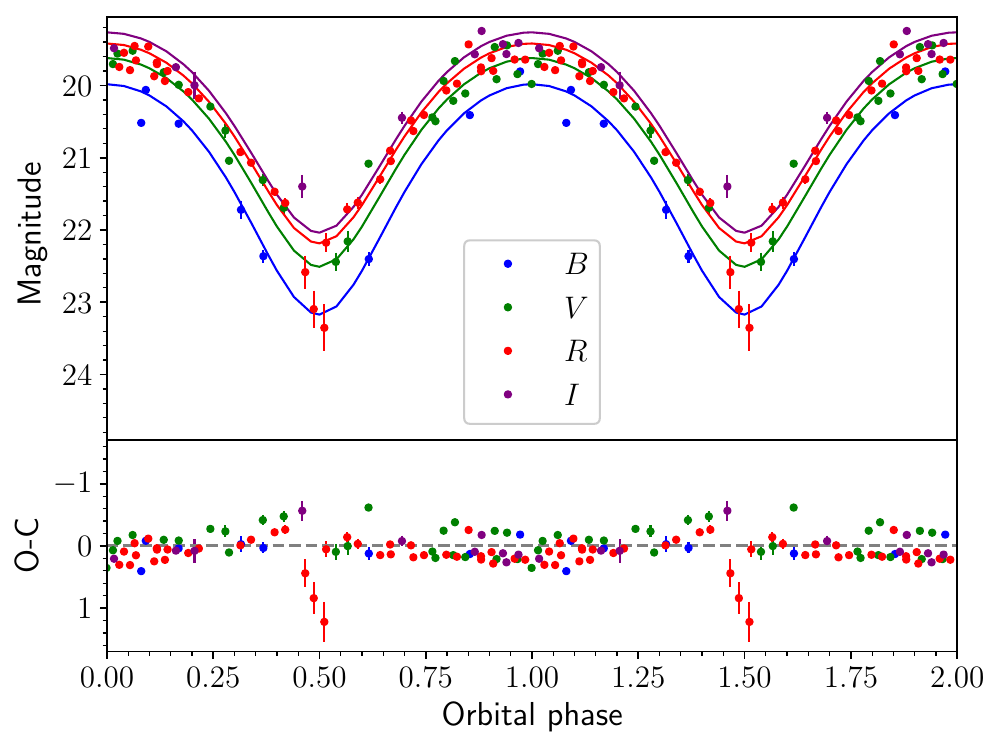}
\includegraphics[width=0.5\linewidth,clip]{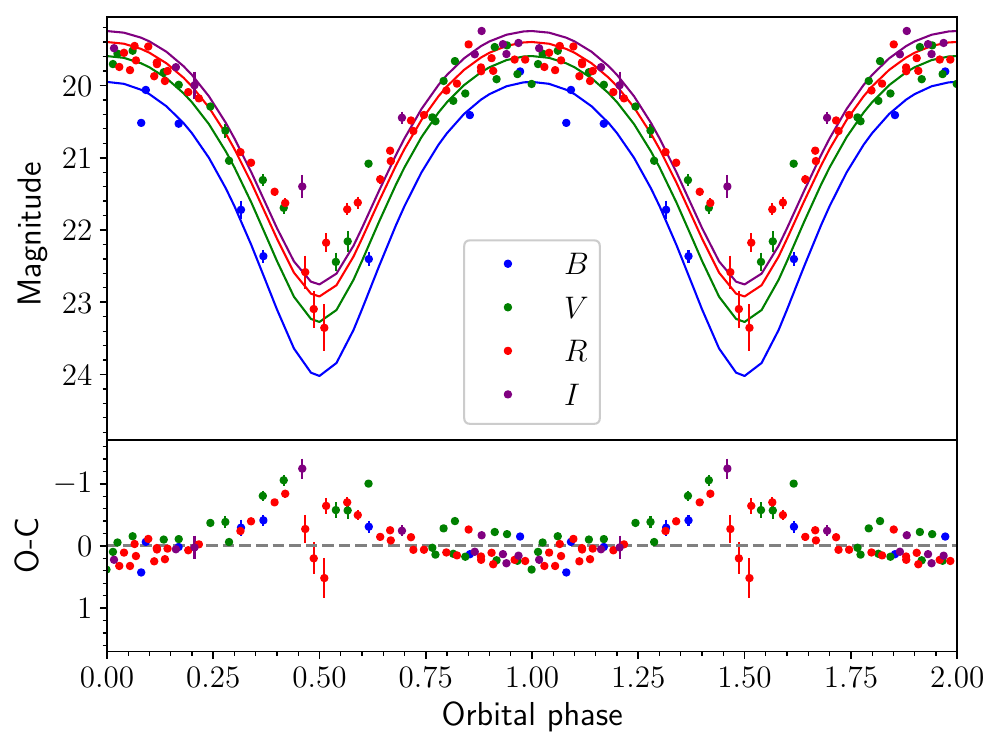}
\end{minipage}
\caption{Folded light curves of \src\ 
obtained in the $B$, $V$, $R$ and $I$ bands with the OAN-SPM 
telescope on July 21--22.
The solid lines show the best-fitting models. In the right panel, the points at the minimum brightness phase in the $R$ band were included in the fit with higher weights (see text). Lower panels show residuals calculated as the difference between the observed (O) and the calculated (C) magnitudes.}
\label{fig:psrLC}
\end{figure*}


The model of the system to fit the light curves considers only two components: the NS as the primary and the heated low-mass companion as the secondary.  
The free parameters were the distance $D$, 
the binary system inclination $i$, the Roche lobe filling factor $f$, 
the `night-side' temperature $T_{\rm n}$ of the secondary, 
the effective irradiation factor $K_{\rm irr}$ [ergs s$^{-1}$ cm$^{-2}$ sr$^{-1}$],  the pulsar mass $M_{\rm NS}$, and mass ratio $q$. 
$K_{\rm irr}$ defines the flux $F_{\rm in}$ transferred to the companion:
\begin{equation}
    F_{\rm in} = \mathrm{cos}(\alpha_{\rm norm}) \Omega \Delta S K_{\rm irr},
\end{equation}
where $\alpha_{\rm norm}$ is
the angle between the normal to the surface and the incoming flux,
$\Omega = \pi R^2_{\rm NS}/a^2$ is the solid angle from which the NS 
is visible from the surface element $\Delta S$ of the secondary, 
$R_{\rm NS}$ is the NS radius and $a$ is the orbit separation.
The `day-side' temperature of the surface element of the secondary is 
\begin{equation}
    T_{\rm d} = T_{\rm n}  \left[1 + \frac{F_{\rm in}}{\sigma_{\rm SB} \Delta S T_{\rm n}^4} \right]^{1/4},
\end{equation}
where $\sigma_{\rm SB}$ is the Stefan-Boltzmann constant.
 We used the gradient descent method to find the minimum of the $\chi^2$ function:
\begin{equation}
\chi^2 = \sum_j^{B,V,R,I}\sum_k^{N} \left(\frac{O_k-C_k}{\sigma_k}\right)^2,
\end{equation}
where $O_k$, $C_k$, and $\sigma_k$ are the observed and the calculated magnitudes, and  
uncertainties 
of the observed magnitudes, respectively.
 We found that the pulsar mass is derived from the fit  with  a very wide uncertainty  of (1--2.7)~\msun\, making the calculated light curves  and the rest of the parameters insensitive to  variations of the mass within this range. 
 We thus fixed it at a canonical NS mass of  $1.4M_\odot$.

The fit results  
are presented in Table~\ref{tab:fit}
and the left panel of Fig.~\ref{fig:psrLC}. The parameter uncertainties were calculated following the method proposed
by \citet{1976ApJ...208..177L}. As seen from the fit residuals, the calculated light curves are generally  consistent with the data, and the minor outliers are likely related to a flaring activity. They are also compatible with the ZTF data which are noisier and less sensitive as compared to the OAN-SPM data. A noticeable    discrepancy is seen only near the minimum brightness phase ($\phi = 0.50\pm0.05$) in the $R$ band.    
The object is reliably ($S/N \approx 4$) detected there (see Fig.~\ref{fig:Rmaxmin}), but high magnitude uncertainties ($\sigma_k\approx 0.25$ mag) lead to low weights of these points in the fit. This formally results in  overshooting  the brightness of the model  points  by about one magnitude in respect to the observed ones.  

At the same time, these points can be critically important for  convincing   constraints 
of the binary system inclination angle.
To study their impact on this parameter, 
we  artificially assigned higher weights to these points, assuming that they could be  detected 
with $\overline{\sigma_k} = 0.09$ mag, which is close to magnitude uncertainties at nearby orbital phases around the minimum, $\phi \in [0.4, 0.45]$ and  $\phi \in [0.55, 0.6]$. 
The result of the respective fit is presented in the right panel of Fig.~\ref{fig:psrLC}. As seen from the plot, the brightness minimum is better approximated  by the model in this case, while the  nearby orbital phases are fitted much worse as compared to the initial fit shown in the left panel.
It is obvious from this experiment that our light curve  model can not account for the brightness dip  at  the minimum of the light curve. This may be caused by variability of the source light curve or by some effects unaccounted for in the model. Given the variable nature of the \src\ light curve, more photometric data are needed to choose between these possibilities. 
However, within the uncertainties most of the parameters obtained in this test fit 
remain the same as in the fit without the artificial weights, 
except for the inclination angle whose best fit value tends to increase by about 10 degrees. 

The resulting interstellar colour excess  
is in agreement with the maximum value 
in the J1838 direction $E(B-V)=0.098$ mag following from the dust map of \citet{dustmap2019}.
Because of a high stochastic  variability of the source, we note that  
the results of the fit   provide  
only rough constraints on the system parameters.
Nevertheless, 
they are compatible 
with parameters 
of other spider systems with modelled light curves.


\begin{table}
\renewcommand{\arraystretch}{1.2}
\caption{The light-curve fitting results for \src.}
\label{tab:fit} 
\begin{center}
\begin{tabular}{lc}
\hline
Fitted parameters &  Values  \\
\hline
Pulsar mass $M_{\rm NS}$, \msun  (fixed)                                   &  1.4   \\          
Mass ratio $q$ =  $M_{\rm c}/M_{\rm NS}$                                        & 0.065(15) \\ 
Distance $D$, kpc                                                               &  3.1(2) \\
`Night-side' temperature $T_{\rm n}$, 10$^3$ K                                  & 2.3(7) \\
Inclination $i$, deg                                                            & 36(10) \\ 
Roche lobe filling factor $f$                                                   & 0.60$_{-0.06}^{+0.10}$ \\
Irradiation factor $K_{\rm irr}$, $10^{20}$ erg~cm$^{-2}$~s$^{-1}$~sr$^{-1}$  & 0.16(3)  \\
Extinction $E(B-V)$, mag  & 0.092(45) \\
\hline
Derived parameters &  \\
\hline
Companion mass $M_{\rm c}$, \msun                                   & 0.10(5) \\
Companion radius $R_{\rm c}$, \rsun                                 & 0.16(3)    \\
`Day-side' temperature: &  \\
Minimum   $T_{\rm d}^{\rm min}$, $10^3$ K              & 6.0(2)   \\
Maximum  $T_{\rm d}^{\rm max}$,  $10^3$ K              & 11.3(4) \\
\hline
\end{tabular}
\end{center}
\end{table}


\section{X-ray and UV data}
\label{sec:x-ray}

We re-analysed the X-ray data from \sw/XRT and UVOT  
together with the extended ROentgen Survey with an Imaging Telescope Array
(\eros; \citealt{erosita2021}) aboard Spectrum-Roentgen-Gamma
(SRG) orbital observatory \citep{Sunyaev2021}.
\sw/XRT observed \src\ nine times in 2019
with the total exposure time of $\approx7.6$ ks.
SRG/\eros\ observed the source field in the course of four all-sky
surveys in 2020--2021 with the total exposure time of $\approx$1.3~ks 
(the vignetting corrected exposure is $\approx$0.7~ks).

Inspection of the \sw\ data revealed that the 
source is clearly seen only
in one data set obtained on 2019 December 12 
with the exposure time of 0.85 ks, between the observations on December 8 and 15 (see the top panels of Fig.~\ref{fig:flare}). 
Thus, the source exhibited a strong flare in X-rays.
A similar flaring activity is simultaneously   observed in the UV with the UVOT 
(see the bottom panels of Fig.~\ref{fig:flare}). We   
measured the AB magnitude during the flare in the $uvw2$ band 
and obtained 20.24(10) mag.
Given $E(B-V) = 0.092$ mag (Table~\ref{tab:fit}), the unabsorbed flux is about 
70~$\mu$Jy. We note that the position of the UV source perfectly coincides with that of the presumed optical counterpart (Fig.~\ref{fig:flare}) 
showing  that we see the same source in the optical, UV and X-rays.

The X-ray spectrum of the source during the flare was extracted 
using the \sw-XRT data products generator
and fitted with the X-Ray Spectral Fitting Package ({\sc xspec}) v.12.11.1 \citep{xspec} using an absorbed power-law (PL) model. 
Only about 30 source counts were obtained in the 0.3--10 keV range.
The spectrum was grouped to ensure at least one count per energy bin.
For the interstellar absorption, we used the {\sc tbabs} model
with the {\sc wilm} abundances \citep*{wilms2000}.
Using the relation from \citet{foight2016}, we transformed  
$E(B-V)$ from the optical light curve fit to the absorbing column density, 
$N_{\rm H}=8\times10^{20}$~cm$^{-2}$, 
which was fixed during the fitting procedure.
The photon index is $\Gamma^{\rm flare}=0.5\pm0.3$ and the unabsorbed flux in the 0.5--10 keV range is 
$F_X^{\rm flare}=4.5^{+1.5}_{-1.1}\times 10^{-12}$ \flux\
(errors correspond to the 1$\sigma$ confidence interval). 
However, we created the light curve using the online generator and
found that the flare took up only about a half of the total exposure (see Fig.~\ref{fig:xrt-lc}).
Thus, the actual flux of the flare 
is about two times greater than the one estimated above, $F_X^{\rm flare}\sim 10^{-11}$ \flux. 

In the \eros\ data \src\ is detected with $\approx$4$\sigma$  significance in the 2.3--8 keV band while it is not seen in the soft band (see Fig.~\ref{fig:srg-img}).
Investigating the individual \eros\ scans, we found that the source is clearly
visible only in the second observation carried out in October 2020  
indicating another hard flare (see Fig.~\ref{fig:srg-lc}).
We extracted the \src\ spectra from each scan using the \eros\ Science Analysis Software System (eSASS). 
They were binned to ensure at least one count per energy bin.
We fitted the spectrum from the second scan with the exposure time of 0.26 ks which contains about 9 counts from the source with the absorbed PL model in the 0.3--10 keV band. 
$N_{\rm H}$ was fixed at the value mentioned above. 
We obtained {a negative best fit vallue of the photon index $\Gamma^{\rm flare}= -1.9^{+0.7}_{-1.0}$ in accord with the fact that the source was detected only in the hard band. The source flux is $F_X^{\rm flare}=9^{+9}_{-5}\times 10^{-12}$~\flux.
Combining spectra from the other three scans, we derived 90 per cent confidence upper limits on the source flux in quiescence, $F_X\lesssim 9\times 10^{-14}$~\flux\ for $\Gamma=1.4$ and $F_X\lesssim 3\times 10^{-14}$~\flux\ for $\Gamma=2.5$. The adopted photon indices are the average values found for BWs and RBs \citep{swihart2022}.

\begin{figure}
\begin{center}
    \includegraphics[width=0.515\linewidth, trim={0 1.3cm 0.8cm 1.3cm},clip]{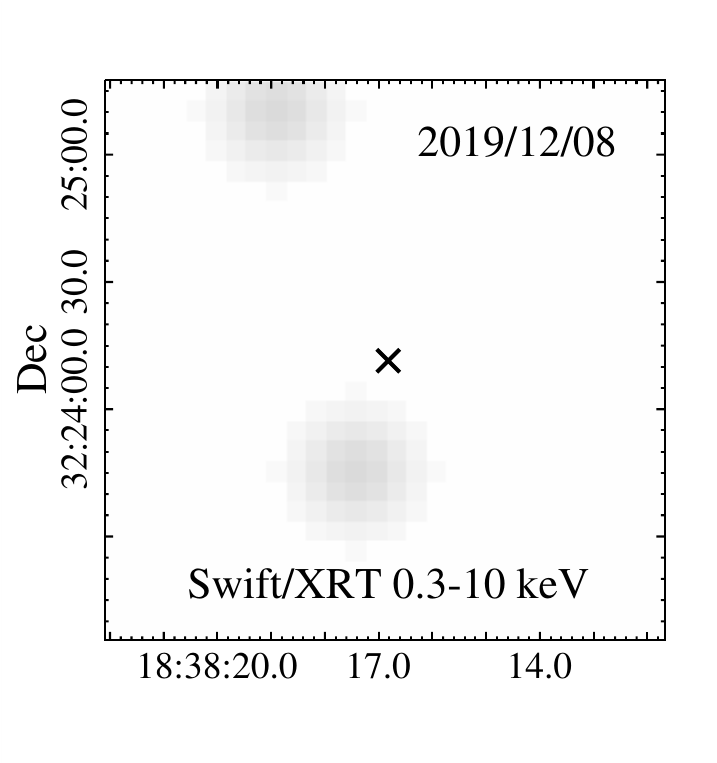}
    \includegraphics[width=0.475\linewidth, trim={1cm 1.3cm 0.8cm 1.3cm},clip]{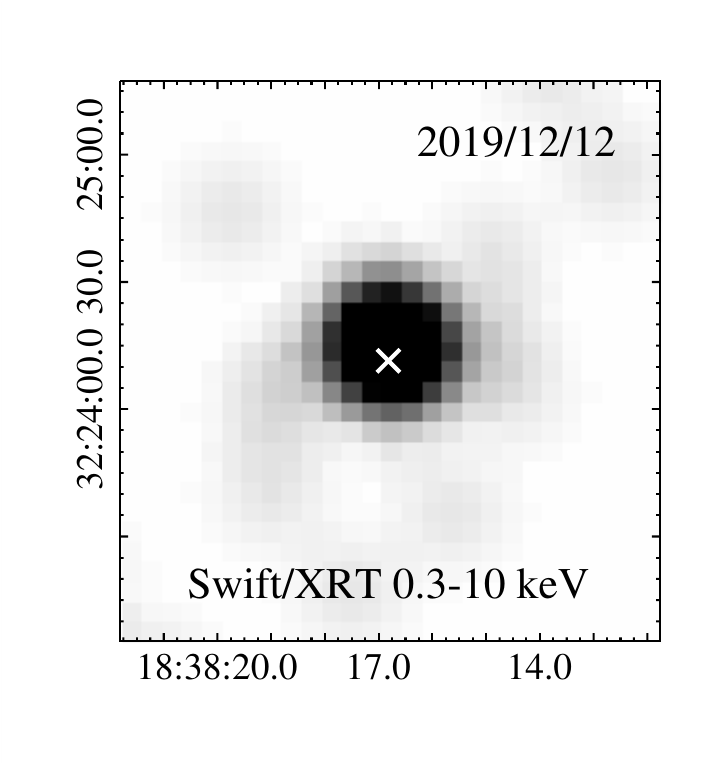}
    \includegraphics[width=0.515\linewidth, trim={0.15cm 0.6cm 0.75cm 1.2cm},clip]{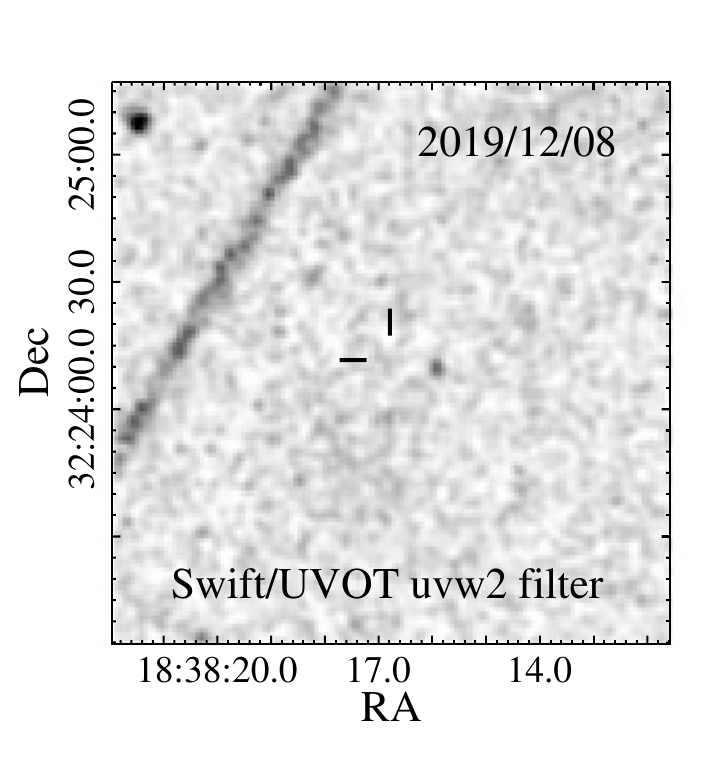}
    \includegraphics[width=0.475\linewidth, trim={1cm 0.6cm 0.8cm 1.2cm},clip]{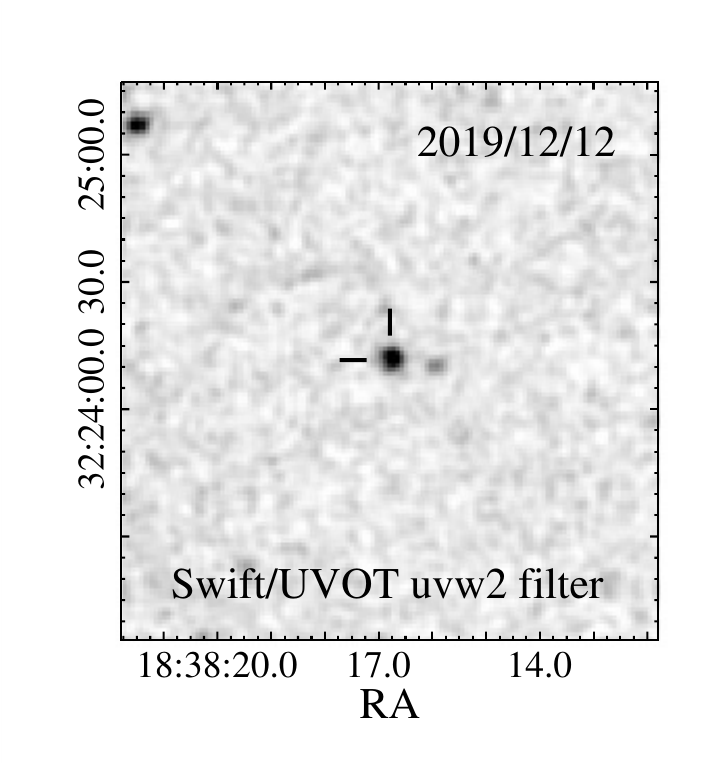}
\end{center}
\caption{\sw\ XRT (top panels) and UVOT (bottom panels) images of the \src\ field.
The position of the \src\ likely optical counterpart is marked by the `X' symbols
in the top panels and by the bars in the bottom panels.
The bright flare occurred on 2019 December 12.  
Exposure times are 0.55 ks for the left panels and 0.85 ks for the right panels.
}
  \label{fig:flare}
\end{figure}

\begin{figure}
 \begin{center}
  \includegraphics[width=1\linewidth,trim={0 0 0 0.3cm},clip]{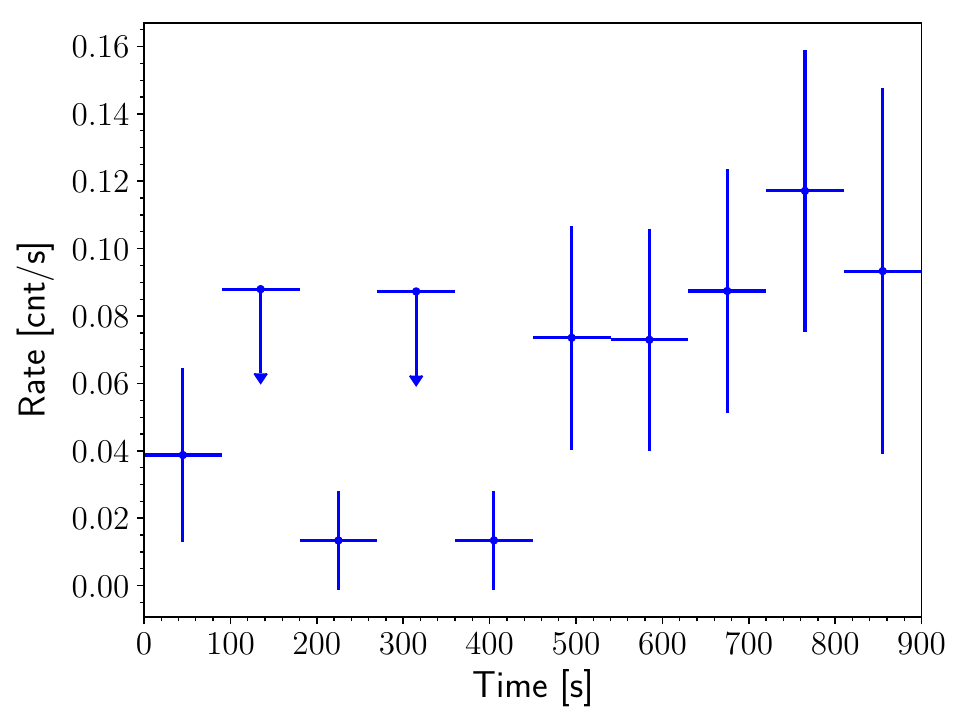}
 \end{center}
 \caption{\sw\ XRT light curve of the \src\ putative counterpart 
 during the flare in the 0.3--10 keV band.
 The time bin size is 90 s. 
 Upper limits correspond to 3$\sigma$ confidence levels.
  }
  \label{fig:xrt-lc}
\end{figure}

\begin{figure}
\begin{center}
    \includegraphics[width=0.51\linewidth, trim={0.7cm 1.2cm 1.7cm 1.1cm},clip]{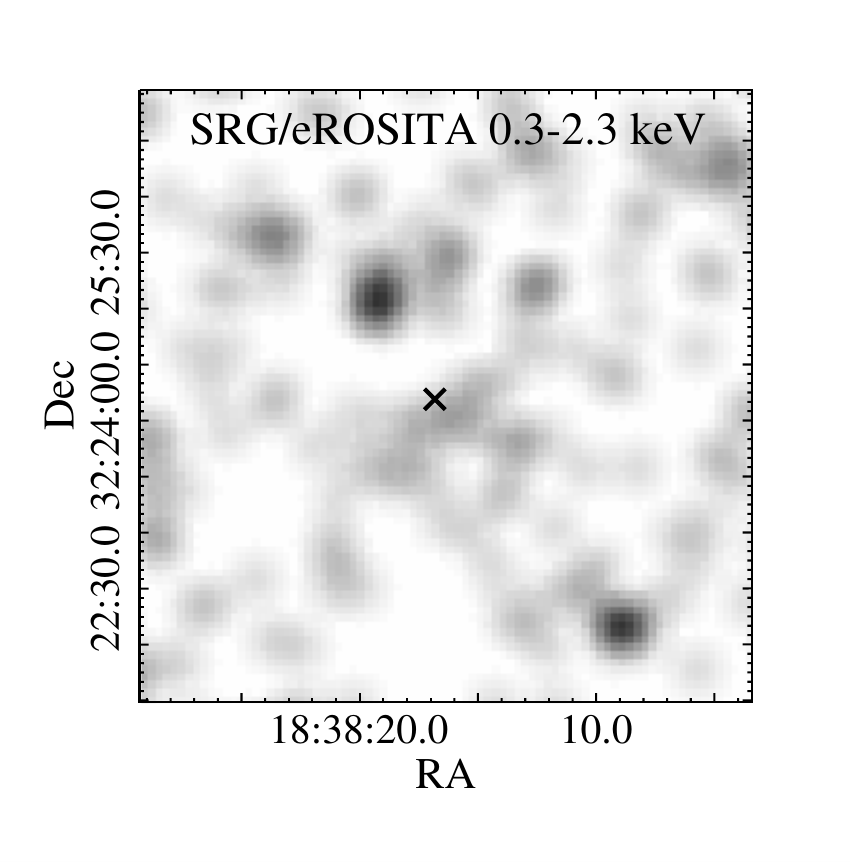}
    \includegraphics[width=0.48\linewidth, trim={1.4cm 1.2cm 1.7cm 1.1cm},clip]{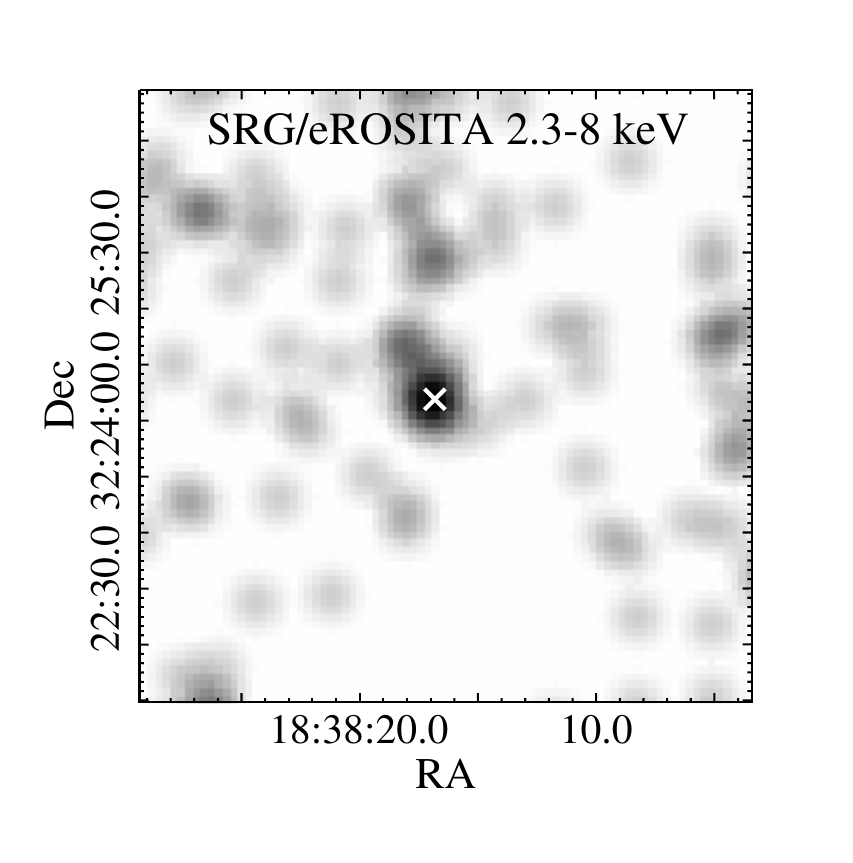}
\end{center}
\caption{SRG/\eros\ images of the \src\ field in the 0.3-2.3 (left) and 2.3-8 (right) keV ranges. The data from the four scans are combined.
The position of the \src\ likely optical counterpart is marked by the `X' symbols.
}
  \label{fig:srg-img}
\end{figure}

\begin{figure}
\begin{center}
    \includegraphics[width=1\linewidth, clip]{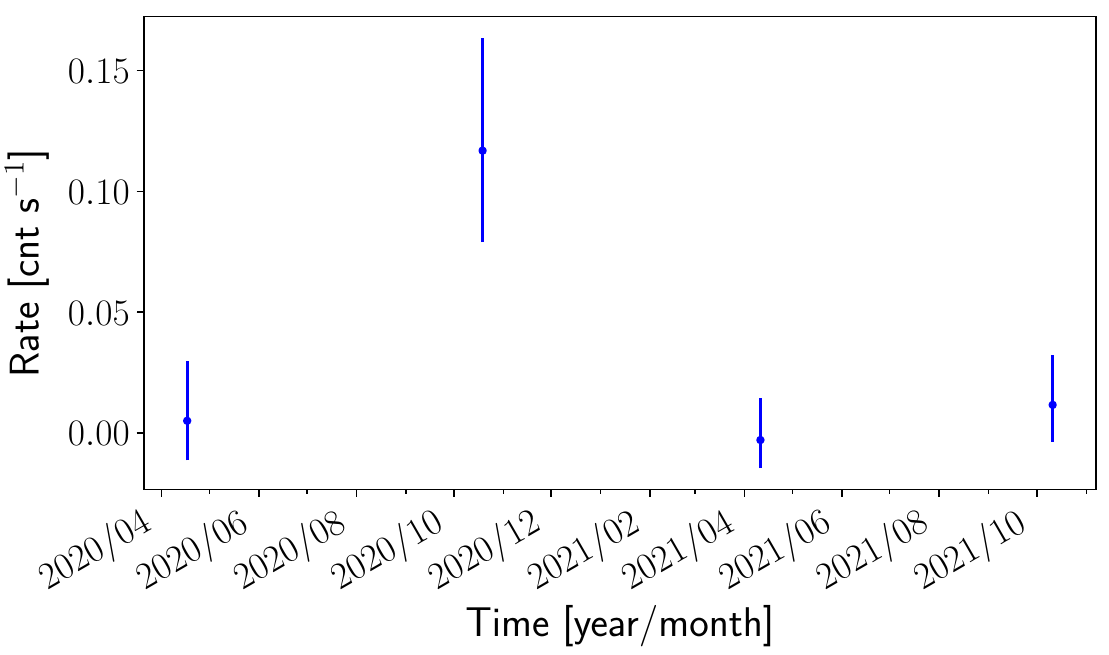}
\end{center}
\caption{SRG \eros\ count rates of the \src\ putative counterpart in individual scans in the 2.3--8 keV band.
The bright flare occurred in October 2020.
}
  \label{fig:srg-lc}
\end{figure}


\section{Discussion}
\label{sec:discussion}

Surface density of X-ray sources within the \sw/XRT field of view with mean fluxes equal to or higher 
than  the mean flux of the  counterpart candidate is  0.0018 per arcmin$^2$. Given that, the chance  probability to find 
an unrelated X-ray source within the 95 per cent (4.1$\times$3.9 arcmin$^2$) 
error ellipse of the $\gamma$-ray source  is  about 8 per cent.
Such probability can be even lower 
if we take into account 
the strong variability of our source. 
\citet{eROSITA-var-sources-2022} found    
only 1325 objects (per a half of sky)   
from which the flux in the 0.3–2.3 keV energy band changed 
by more than a factor of 10 during the SRG/\eros\ all-sky survey. 
The reported number transforms to about 1.8$\times$10$^{-5}$ 
variable  objects per arcmin$^2$ and the chance probability of 0.1 per cent. 
This supports the association of the X-ray source and the  
optical object  with the pulsar candidate proposed 
by \citet{Kerby2021}. However, this probability estimation    
has to be considered with a caution, 
as the local density of variable sources can be different from a mean one, and it may depend on their maximum to minimum flux ratio. 

Most RBs have light curves with two peaks per orbital cycle 
caused by tidal distortion of the companion, i.e. ellipsoidal variations. 
For BWs, irradiation from the pulsar wind is more significant 
than the tidal distortion effects which leads to the heating 
of the companion's side facing  the pulsar and a single-peak
light curve \citep[e.g.][]{draghis2019,swihart2022}. Usual light curve 
peak-to-peak amplitude is $\lesssim$1~mag for RBs and 2--4 mag for BWs. 

We have revealed that the 
optical and X-ray counterpart of \src\
proposed by \citet{Kerby2021}
is a highly variable source demonstrating typical properties of spider systems. The presumed optical counterpart demonstrates brightness changes with a period of
$\approx$4~h, resembling  an orbital period of the expected spider system. 
It shows a single-peaked sine-like variability with a total amplitude of the brightness variation of about 3 mag (from
$\approx$20 to $\approx$23 mag in the $R$ band). At the same time, we found indications of its  strong 
short-term flaring activity.
The shape and amplitudes of the  light curves are more 
typical for BWs. However, the RB interpretation of \src\ cannot be 
excluded since there is an example of a RB pulsar, J2339$-$0533, with an orbital period of 4.6~h, 
which also demonstrates single-peak light curves with a large 
difference between maxima and minima (e.g. $\Delta g\approx 6$ mag, 
\citealt{kandel2020}). It was assumed to belong to the BW
subclass until the pulsar discovery and measurements of 
the binary parameters \citep{ray2020}. 

From our light curve modelling of \src, the derived Roche lobe filling factor is $f\approx 0.6$. 
It excludes any significant optical brightness  variations  related with the distorted shape of the companion.
Nevertheless, it is also less than what is usually found for most spider systems, 
although there are objects with low values, e.g., PSR J0023+0923 with $f$ = 0.5$\pm$0.1  \citep{matasanchez2023}.

Our modelling also showed that  
\src\ demonstrates a significant difference between the night-side ($T_{\rm n}$ $\sim 2300$ K, corresponding to  a M5--L4 type star) and day-side temperatures ($T_{\rm d}$ $\sim 11000$ K). 
RBs typically have $T_{\rm n}$~=~4000--6000 K 
\citep[e.g.][]{PSR1622arXiv}  
whereas BWs night sides are cooler, with $T_{\rm n}$~=~1000--3000 K \citep[e.g.][]{matasanchez2023}. 
In addition, RBs with their larger orbital periods usually have lower irradiation temperatures\footnote{The day-side temperature is related to the night-side and irradiation temperatures as $T_{\rm d}^4=T_{\rm n}^4 + T_{\rm irr}^4$.}
$T_{\rm irr}$ (several hundreds K) in comparison with BWs ($\gtrsim 3000$ K). 
A large $T_{\rm irr}$ of $\approx6300$~K was found for the above mentioned RB J2339$-$0533 with extreme heating \citep{kandel2020}. 
However, this is still lower than what we obtained for \src\ 
($T_{\rm irr} \approx T_{\rm d}^{\rm max} = 10600$ K). 
The companion of the RB PSR J1816+4510 has an unusually high temperature of 16000 K \citep{kaplan2013} but 
its variability over the orbit is low \citep[$\Delta G\sim0.1$~mag;][]{Koljonen&Linares2023}. 
There are several BW systems with $T_{\rm irr}$ of about 9000--10000 K,
PSRs J1810+1744, J1555$-$2908 and J1641+8049 \citep{romani2021,kennedy2022,matasanchez2023},
as well as the BW candidate ZTF J1406+1222 \citep{burge2022}.
Overall, the temperature distribution of the \src\ companion is more typical for the BW family, although it is necessary to obtain higher resolution light curves to confirm this.

Rapid flaring in the optical revealed for \src\ 
is  observed for some RB and BW systems:
BW PSR J1311$-$3430 \citep{romain2012},
BW candidates 4FGL~J0935.3+0901 \citep{Flaring-BW-2022}
and 4FGL~J1408.6$-$2917 \citep{swihart2022}, 
RB PSR J1048+2339, RB candidates XMMU J083850.38$-$282756.8 
and 1FGL J0523.5$-$2529 \citep*{Flaring2018,Flaring-RB-2022}.
The brightest ($\Delta$ mag $\approx$1) optical flare 
of \src\ was detected on 2022 July 24
(Fig.~\ref{fig:lc-BVRI}). 
The origin of such flares is not clear. They can be attributed
to the variable emission from the intro-binary shocks between the pulsar and the companion created by the interaction of the pulsar wind and the wind ejected from the companion.
Otherwise, they can be caused by an intrinsic companion magnetic activity as observed for low-mass stars \citep[e.g.][]{Schmidt2019}.

The mass distribution of spider pulsars is bimodal with a gap in the range 
0.07--0.1~\msun\ 
between BWs and RBs \citep{swihart2022}. Thus, \src\ with $M_{\rm c}=0.10(5)$~\msun\  
can belong to any of these subclasses. 
Better constraints on $M_{\rm c}$ are necessary  
to understand  its spider type. 

The distance obtained from the light curve modelling is about 3.1 kpc which is close to the upper bound of the distance range provided by \gaia\ (Table~\ref{tab:pars}). 
The transverse velocity corresponding to this distance is about 200 km~s$^{-1}$ which is at the upper bound of the velocity distribution for binary pulsar systems \citep{hobbs2005}.

In X-rays, \src\ was firmly detected  during two strong flares.
Thus, we can estimate only an upper limit on its luminosity
in quiescence, $L_X \lesssim 3.5\times10^{31} (D/3.1\ {\rm kpc})^2$ \ergs\ and
$\lesssim 10^{32} (D/3.1\ {\rm kpc})^2$ \ergs\ 
for the PL model with the photon index appropriate for BWs and RBs, respectively.
Studying X-ray properties of spider pulsars, \citet{swihart2022} found that 
RBs are brighter than BWs with the average luminosities 
$L_X^{\rm rb}=2.3\times10^{32}$ \ergs\ and 
$L_X^{\rm bw}=1.4\times10^{31}$ \ergs. 
The scatters of luminosities are rather wide and 
\src\ upper limits can be consistent with both subclasses.

Strong X-ray flares were observed for PSR~J1311$-$3430,
PSR~J1048+2339, XMMU J083850.38$-$282756.8 
and 1FGL J0523.5$-$2529 systems mentioned
above. The brightest flares 
were detected for 1FGL J0523.5$-$2529 with the peak luminosities
of $\sim10^{34}$~\ergs, which is a factor of $\sim$100 stronger than its  minimum luminosity 
\citep{Flaring-RB-2022}. The X-ray flares of \src\ with the luminosity  
$L_X\sim 10^{34}(D/{\rm 3.1\ kpc})^{2}$ \ergs\ are at least
two orders of magnitude 
(by a factor of 100--200) 
brighter than its quiescent emission. 
Its flare luminosity can be as large as for 
1FGL J0523.5$-$2529, if the distance to the source is indeed 3.1 kpc. 
Thus, its flares can be the most luminous among the 
observed in RB/BW systems without accretion disks. 
The spectra of both flares are hard. 
It is interesting that the flare detected with \eros\ likely demonstrates 
a negative photon index.
This possibly can be explained by an unresolved time lag  between the hard and soft photon emission with a brighter outburst in the hard band.


 Any alternative  interpretation of 
the data, such as a   cataclysmic 
variable (CV), a tight binary stellar system or  a rotating magnetic star, appears to be very unlikely mainly due to the high modulation  of the measured light curves.  
The former two demonstrate much 
flatter 
orbital light curves, sometimes with sharp dips due to binary companion eclipses \citep[e.g.,][and references therein]{warner1995,zasche2017}. 
For instance, light curves of intermediate 
polars (IPs), a subclass of CVs,  
typically show about 0.5 mag 
amplitude periodic variability. 
Sharp dips of 1--2 mag were observed
in a few eclipsing IPs (e.g.: V597 Pup, \citet{2009MNRAS.397..979W}; IPHAS J062746.41+014811.3, \citet{2012ApJ...758...79A}; CXOGBS J174954.5-294335, \citet{2017MNRAS.466..129J}).
Brightness variations of magnetic stars are typically less than  a magnitude while rotational periods are greater than  a day \citep[e.g.,][]{2021MNRAS}.    
Therefore, the spider pulsar interpretation  currently remains the most plausible. This is also strongly supported by the detection of  $\gamma$-rays which are not observed in the alternative cases. 
The fact that the estimated $\gamma$-ray luminosity of J1838
(Table~\ref{tab:pars}) is within the range 
of luminosities ($4\times10^{32}$ -- $4\times10^{34}$ \ergs) of the spider pulsar family \citep{strader2019,swihart2022}
is in line with this interpretation. 
 
Optical spectroscopy is needed to obtain the radial velocity curve of the \src\ companion candidate and confirm its nature. Accounting for its brightness, this is feasible 
with 8--10 meter class telescopes. 
RBs spectra are dominated by hydrogen lines while BWs spectra may 
show helium features indicating hydrogen-deficient atmospheres (\citealt{swihart2022} and references therein).
Measurements of the companion's radial velocity would provide constraints 
on the mass of the putative pulsar.
High time-resolution multi-band light curves, especially in the minimum phase, would allow one to constrain the fundamental parameters of the binary system with a higher precision. 
To better understand the mechanism
of flares simultaneous multicolour photometry would be preferable. Deeper X-ray observations are necessary to detect the object in quiescence. Last but not least, J1838 is of a particular interest for targeted searches of millisecond pulsations in the 
radio and/or $\gamma$-ray domains. 

}

\section*{Acknowledgements}

Based upon observations carried out at the Observatorio Astronómico Nacional on the Sierra San Pedro Mártir (OAN-SPM), Baja California, México.
We thank the daytime and night support staff at the OAN-SPM for facilitating and helping obtain our observations.
This work has made use of data from the European Space Agency (ESA) mission
{\it Gaia} (\url{https://www.cosmos.esa.int/gaia}), processed by the {\it Gaia}
Data Processing and Analysis Consortium (DPAC,
\url{https://www.cosmos.esa.int/web/gaia/dpac/consortium}).
Funding for the DPAC
has been provided by national institutions, in particular the institutions
participating in the {\it Gaia} Multilateral Agreement.
Based on observations obtained with the Samuel Oschin Telescope 48-inch and the 60-inch Telescope at the Palomar Observatory as part of the Zwicky Transient Facility project. ZTF is supported by the National Science Foundation under Grants No. AST-1440341 and AST-2034437 and a collaboration including current partners Caltech, IPAC, the Weizmann Institute for
Science, the Oskar Klein Center at Stockholm University, the University of Maryland, Deutsches Elektronen-Synchrotron and Humboldt University, the TANGO Consortium of Taiwan, the University of Wisconsin at Milwaukee, Trinity College Dublin, Lawrence Livermore National Laboratories, IN2P3, University of Warwick, Ruhr University Bochum, Northwestern University and former partners the University of Washington, Los Alamos National Laboratories, and Lawrence Berkeley National Laboratories. Operations are conducted by COO, IPAC, and UW.
This work used data obtained with eROSITA telescope onboard SRG observatory. The SRG observatory was built by Roskosmos in the interests of the Russian Academy of Sciences represented by its Space Research Institute (IKI) in the framework of the Russian Federal Space Program, with the participation of the Deutsches Zentrum für Luft- und Raumfahrt (DLR). The SRG/eROSITA X-ray telescope was built by a consortium of German Institutes led by MPE, and supported by DLR.  The SRG spacecraft was designed, built, launched and is operated by the Lavochkin Association and its subcontractors. The science data are downlinked via the Deep Space Network Antennae in Bear Lakes, Ussurijsk, and Baykonur, funded by Roskosmos. The eROSITA data used in this work were processed using the eSASS software system developed by the German eROSITA consortium and proprietary data reduction and analysis software developed by the Russian eROSITA Consortium.

 We are grateful to the anonymous referee for useful comments allowing us to improve the paper.
The work of DAZ and AVK was supported by the Russian Science Foundation, 
grant number 22-22-00921, \url{https://rscf.ru/project/22-22-00921/}.
DAZ thanks Pirinem School of Theoretical Physics for hospitality. 
SVZ acknowledges PAPIIT grant IN119323.
\section*{Data Availability}

The \sw\ data are available through the archive
\url{https://www.swift.ac.uk/swift_portal/}, ZTF data --  \url{https://irsa.ipac.caltech.edu/Missions/ztf.html}, \eros\ and OAN-SPM data -- on request.



\bibliographystyle{mnras}
\bibliography{ref} 

\begin{thebibliography}{}
\makeatletter
\relax
\def\mn@urlcharsother{\let\do\@makeother \do\$\do\&\do\#\do\^\do\_\do\%\do\~}
\def\mn@doi{\begingroup\mn@urlcharsother \@ifnextchar [ {\mn@doi@}
  {\mn@doi@[]}}
\def\mn@doi@[#1]#2{\def\@tempa{#1}\ifx\@tempa\@empty \href
  {http://dx.doi.org/#2} {doi:#2}\else \href {http://dx.doi.org/#2} {#1}\fi
  \endgroup}
\def\mn@eprint#1#2{\mn@eprint@#1:#2::\@nil}
\def\mn@eprint@arXiv#1{\href {http://arxiv.org/abs/#1} {{\tt arXiv:#1}}}
\def\mn@eprint@dblp#1{\href {http://dblp.uni-trier.de/rec/bibtex/#1.xml}
  {dblp:#1}}
\def\mn@eprint@#1:#2:#3:#4\@nil{\def\@tempa {#1}\def\@tempb {#2}\def\@tempc
  {#3}\ifx \@tempc \@empty \let \@tempc \@tempb \let \@tempb \@tempa \fi \ifx
  \@tempb \@empty \def\@tempb {arXiv}\fi \@ifundefined
  {mn@eprint@\@tempb}{\@tempb:\@tempc}{\expandafter \expandafter \csname
  mn@eprint@\@tempb\endcsname \expandafter{\@tempc}}}

\bibitem[\protect\citeauthoryear{{Abdollahi} et~al.,}{{Abdollahi}
  et~al.}{2022}]{4fgl-dr3}
{Abdollahi} S.,  et~al., 2022, \mn@doi [\apjs] {10.3847/1538-4365/ac6751},
  \href {https://ui.adsabs.harvard.edu/abs/2022ApJS..260...53A} {260, 53}

\bibitem[\protect\citeauthoryear{{Ablimit}}{{Ablimit}}{2019}]{ablimit2019}
{Ablimit} I.,  2019, \mn@doi [\apj] {10.3847/1538-4357/ab339d}, \href
  {https://ui.adsabs.harvard.edu/abs/2019ApJ...881...72A} {881, 72}

\bibitem[\protect\citeauthoryear{{Archibald} et~al.,}{{Archibald}
  et~al.}{2009}]{archibald2009}
{Archibald} A.~M.,  et~al., 2009, \mn@doi [Science] {10.1126/science.1172740},
  \href {https://ui.adsabs.harvard.edu/abs/2009Sci...324.1411A} {324, 1411}

\bibitem[\protect\citeauthoryear{{Arnaud}}{{Arnaud}}{1996}]{xspec}
{Arnaud} K.~A.,  1996, in {Jacoby} G.~H.,  {Barnes} J.,  eds,  Astronomical
  Society of the Pacific Conference Series Vol. 101, Astronomical Data Analysis
  Software and Systems V. p.~17

\bibitem[\protect\citeauthoryear{{Au} et~al.,}{{Au} et~al.}{2023}]{au2023}
{Au} K.-Y.,  et~al., 2023, \mn@doi [\apj] {10.3847/1538-4357/acae8a}, \href
  {https://ui.adsabs.harvard.edu/abs/2023ApJ...943..103A} {943, 103}

\bibitem[\protect\citeauthoryear{{Aungwerojwit}, {G{\"a}nsicke}, {Wheatley},
  {Pyrzas}, {Staels}, {Krajci}  \& {Rodr{\'\i}guez-Gil}}{{Aungwerojwit}
  et~al.}{2012}]{2012ApJ...758...79A}
{Aungwerojwit} A.,  {G{\"a}nsicke} B.~T.,  {Wheatley} P.~J.,  {Pyrzas} S.,
  {Staels} B.,  {Krajci} T.,   {Rodr{\'\i}guez-Gil} P.,  2012, \mn@doi [\apj]
  {10.1088/0004-637X/758/2/79}, \href
  {https://ui.adsabs.harvard.edu/abs/2012ApJ...758...79A} {758, 79}

\bibitem[\protect\citeauthoryear{{Bassa} et~al.,}{{Bassa}
  et~al.}{2014}]{bassa2014}
{Bassa} C.~G.,  et~al., 2014, \mn@doi [\mnras] {10.1093/mnras/stu708}, \href
  {https://ui.adsabs.harvard.edu/abs/2014MNRAS.441.1825B} {441, 1825}

\bibitem[\protect\citeauthoryear{{Benvenuto}, {De Vito}  \&
  {Horvath}}{{Benvenuto} et~al.}{2014}]{benvenuto2014}
{Benvenuto} O.~G.,  {De Vito} M.~A.,   {Horvath} J.~E.,  2014, \mn@doi [\apjl]
  {10.1088/2041-8205/786/1/L7}, \href
  {https://ui.adsabs.harvard.edu/abs/2014ApJ...786L...7B} {786, L7}

\bibitem[\protect\citeauthoryear{{Bernhard}, {H{\"u}mmerich}, {Paunzen}  \&
  {Sup{\'\i}kov{\'a}}}{{Bernhard} et~al.}{2021}]{2021MNRAS}
{Bernhard} K.,  {H{\"u}mmerich} S.,  {Paunzen} E.,   {Sup{\'\i}kov{\'a}} J.,
  2021, \mn@doi [\mnras] {10.1093/mnras/stab2065}, \href
  {https://ui.adsabs.harvard.edu/abs/2021MNRAS.506.4561B} {506, 4561}

\bibitem[\protect\citeauthoryear{{Burdge} et~al.,}{{Burdge}
  et~al.}{2022}]{burge2022}
{Burdge} K.~B.,  et~al., 2022, \mn@doi [\nat] {10.1038/s41586-022-04551-1},
  \href {https://ui.adsabs.harvard.edu/abs/2022Natur.605...41B} {605, 41}

\bibitem[\protect\citeauthoryear{{Chen}, {Chen}, {Tauris}  \& {Han}}{{Chen}
  et~al.}{2013}]{chen2013}
{Chen} H.-L.,  {Chen} X.,  {Tauris} T.~M.,   {Han} Z.,  2013, \mn@doi [\apj]
  {10.1088/0004-637X/775/1/27}, \href
  {https://ui.adsabs.harvard.edu/abs/2013ApJ...775...27C} {775, 27}

\bibitem[\protect\citeauthoryear{{Cho}, {Halpern}  \& {Bogdanov}}{{Cho}
  et~al.}{2018}]{Flaring2018}
{Cho} P.~B.,  {Halpern} J.~P.,   {Bogdanov} S.,  2018, \mn@doi [\apj]
  {10.3847/1538-4357/aade92}, \href
  {https://ui.adsabs.harvard.edu/abs/2018ApJ...866...71C} {866, 71}

\bibitem[\protect\citeauthoryear{{Draghis}, {Romani}, {Filippenko}, {Brink},
  {Zheng}, {Halpern}  \& {Camilo}}{{Draghis} et~al.}{2019}]{draghis2019}
{Draghis} P.,  {Romani} R.~W.,  {Filippenko} A.~V.,  {Brink} T.~G.,  {Zheng}
  W.,  {Halpern} J.~P.,   {Camilo} F.,  2019, \mn@doi [\apj]
  {10.3847/1538-4357/ab378b}, \href
  {https://ui.adsabs.harvard.edu/abs/2019ApJ...883..108D} {883, 108}

\bibitem[\protect\citeauthoryear{{Evans} et~al.,}{{Evans}
  et~al.}{2009}]{evans2009}
{Evans} P.~A.,  et~al., 2009, \mn@doi [\mnras]
  {10.1111/j.1365-2966.2009.14913.x}, \href
  {https://ui.adsabs.harvard.edu/abs/2009MNRAS.397.1177E} {397, 1177}

\bibitem[\protect\citeauthoryear{{Flewelling} et~al.,}{{Flewelling}
  et~al.}{2020}]{ps2020}
{Flewelling} H.~A.,  et~al., 2020, \mn@doi [\apjs] {10.3847/1538-4365/abb82d},
  \href {https://ui.adsabs.harvard.edu/abs/2020ApJS..251....7F} {251, 7}

\bibitem[\protect\citeauthoryear{{Foight}, {G{\"u}ver}, {{\"O}zel}  \&
  {Slane}}{{Foight} et~al.}{2016}]{foight2016}
{Foight} D.~R.,  {G{\"u}ver} T.,  {{\"O}zel} F.,   {Slane} P.~O.,  2016,
  \mn@doi [\apj] {10.3847/0004-637X/826/1/66}, \href
  {https://ui.adsabs.harvard.edu/abs/2016ApJ...826...66F} {826, 66}

\bibitem[\protect\citeauthoryear{{Gaia Collaboration} et~al.,}{{Gaia
  Collaboration} et~al.}{2016}]{gaia2016}
{Gaia Collaboration} et~al., 2016, \mn@doi [\aap]
  {10.1051/0004-6361/201629272}, \href
  {https://ui.adsabs.harvard.edu/abs/2016A&A...595A...1G} {595, A1}

\bibitem[\protect\citeauthoryear{{Gaia Collaboration} et~al.,}{{Gaia
  Collaboration} et~al.}{2021}]{gaia2021_edr3}
{Gaia Collaboration} et~al., 2021, \mn@doi [\aap]
  {10.1051/0004-6361/202039657}, \href
  {https://ui.adsabs.harvard.edu/abs/2021A&A...649A...1G} {649, A1}

\bibitem[\protect\citeauthoryear{{Green}, {Schlafly}, {Zucker}, {Speagle}  \&
  {Finkbeiner}}{{Green} et~al.}{2019}]{dustmap2019}
{Green} G.~M.,  {Schlafly} E.,  {Zucker} C.,  {Speagle} J.~S.,   {Finkbeiner}
  D.,  2019, \mn@doi [\apj] {10.3847/1538-4357/ab5362}, \href
  {https://ui.adsabs.harvard.edu/abs/2019ApJ...887...93G} {887, 93}

\bibitem[\protect\citeauthoryear{{Guo}, {Wang}  \& {Han}}{{Guo}
  et~al.}{2022}]{guo2022}
{Guo} Y.,  {Wang} B.,   {Han} Z.,  2022, \mn@doi [\mnras]
  {10.1093/mnras/stac1917}, \href
  {https://ui.adsabs.harvard.edu/abs/2022MNRAS.515.2725G} {515, 2725}

\bibitem[\protect\citeauthoryear{{Halpern}}{{Halpern}}{2022}]{Flaring-BW-2022}
{Halpern} J.~P.,  2022, \mn@doi [\apjl] {10.3847/2041-8213/ac746f}, \href
  {https://ui.adsabs.harvard.edu/abs/2022ApJ...932L...8H} {932, L8}

\bibitem[\protect\citeauthoryear{{Halpern}, {Perez}  \& {Bogdanov}}{{Halpern}
  et~al.}{2022}]{Flaring-RB-2022}
{Halpern} J.~P.,  {Perez} K.~I.,   {Bogdanov} S.,  2022, \mn@doi [\apj]
  {10.3847/1538-4357/ac8161}, \href
  {https://ui.adsabs.harvard.edu/abs/2022ApJ...935..151H} {935, 151}

\bibitem[\protect\citeauthoryear{{Hobbs}, {Lorimer}, {Lyne}  \&
  {Kramer}}{{Hobbs} et~al.}{2005}]{hobbs2005}
{Hobbs} G.,  {Lorimer} D.~R.,  {Lyne} A.~G.,   {Kramer} M.,  2005, \mn@doi
  [\mnras] {10.1111/j.1365-2966.2005.09087.x}, \href
  {https://ui.adsabs.harvard.edu/abs/2005MNRAS.360..974H} {360, 974}

\bibitem[\protect\citeauthoryear{{Johnson} et~al.,}{{Johnson}
  et~al.}{2017}]{2017MNRAS.466..129J}
{Johnson} C.~B.,  et~al., 2017, \mn@doi [\mnras] {10.1093/mnras/stw3063}, \href
  {https://ui.adsabs.harvard.edu/abs/2017MNRAS.466..129J} {466, 129}

\bibitem[\protect\citeauthoryear{{Kandel}, {Romani}, {Filippenko}, {Brink}  \&
  {Zheng}}{{Kandel} et~al.}{2020}]{kandel2020}
{Kandel} D.,  {Romani} R.~W.,  {Filippenko} A.~V.,  {Brink} T.~G.,   {Zheng}
  W.,  2020, \mn@doi [\apj] {10.3847/1538-4357/abb6fd}, \href
  {https://ui.adsabs.harvard.edu/abs/2020ApJ...903...39K} {903, 39}

\bibitem[\protect\citeauthoryear{{Kaplan}, {Bhalerao}, {van Kerkwijk},
  {Koester}, {Kulkarni}  \& {Stovall}}{{Kaplan} et~al.}{2013}]{kaplan2013}
{Kaplan} D.~L.,  {Bhalerao} V.~B.,  {van Kerkwijk} M.~H.,  {Koester} D.,
  {Kulkarni} S.~R.,   {Stovall} K.,  2013, \mn@doi [\apj]
  {10.1088/0004-637X/765/2/158}, \href
  {https://ui.adsabs.harvard.edu/abs/2013ApJ...765..158K} {765, 158}

\bibitem[\protect\citeauthoryear{{Karpova}, {Zyuzin}, {Shibanov}  \&
  {Gilfanov}}{{Karpova} et~al.}{2023}]{karpova2023}
{Karpova} A.~V.,  {Zyuzin} D.~A.,  {Shibanov} Y.~A.,   {Gilfanov} M.~R.,  2023,
  \mn@doi [\mnras] {10.1093/mnras/stad1992}, \href
  {https://ui.adsabs.harvard.edu/abs/2023MNRAS.524.3020K} {524, 3020}

\bibitem[\protect\citeauthoryear{{Kennedy} et~al.,}{{Kennedy}
  et~al.}{2022}]{kennedy2022}
{Kennedy} M.~R.,  et~al., 2022, \mn@doi [\mnras] {10.1093/mnras/stac379}, \href
  {https://ui.adsabs.harvard.edu/abs/2022MNRAS.512.3001K} {512, 3001}

\bibitem[\protect\citeauthoryear{{Kerby} et~al.,}{{Kerby}
  et~al.}{2021}]{Kerby2021}
{Kerby} S.,  et~al., 2021, \mn@doi [\apj] {10.3847/1538-4357/ac2e91}, \href
  {https://ui.adsabs.harvard.edu/abs/2021ApJ...923...75K} {923, 75}

\bibitem[\protect\citeauthoryear{{Koljonen} \& {Linares}}{{Koljonen} \&
  {Linares}}{2023}]{Koljonen&Linares2023}
{Koljonen} K. I.~I.,  {Linares} M.,  2023, arXiv e-prints, \href
  {https://ui.adsabs.harvard.edu/abs/2023arXiv230807377K} {p. arXiv:2308.07377}

\bibitem[\protect\citeauthoryear{{Lampton}, {Margon}  \& {Bowyer}}{{Lampton}
  et~al.}{1976}]{1976ApJ...208..177L}
{Lampton} M.,  {Margon} B.,   {Bowyer} S.,  1976, \mn@doi [\apj]
  {10.1086/154592}, \href
  {https://ui.adsabs.harvard.edu/abs/1976ApJ...208..177L} {208, 177}

\bibitem[\protect\citeauthoryear{{Landolt}}{{Landolt}}{1992}]{1992AJ....104..340L}
{Landolt} A.~U.,  1992, \mn@doi [\aj] {10.1086/116242}, \href
  {https://ui.adsabs.harvard.edu/abs/1992AJ....104..340L} {104, 340}

\bibitem[\protect\citeauthoryear{{Linares}, {Shahbaz}  \& {Casares}}{{Linares}
  et~al.}{2018}]{linares2018}
{Linares} M.,  {Shahbaz} T.,   {Casares} J.,  2018, \mn@doi [\apj]
  {10.3847/1538-4357/aabde6}, \href
  {https://ui.adsabs.harvard.edu/abs/2018ApJ...859...54L} {859, 54}

\bibitem[\protect\citeauthoryear{{Lindegren} et~al.,}{{Lindegren}
  et~al.}{2021}]{gaia-edr3-astrometry}
{Lindegren} L.,  et~al., 2021, \mn@doi [\aap] {10.1051/0004-6361/202039709},
  \href {https://ui.adsabs.harvard.edu/abs/2021A&A...649A...2L} {649, A2}

\bibitem[\protect\citeauthoryear{{Lomb}}{{Lomb}}{1976}]{lomb1976}
{Lomb} N.~R.,  1976, \mn@doi [\apss] {10.1007/BF00648343}, \href
  {https://ui.adsabs.harvard.edu/abs/1976Ap&SS..39..447L} {39, 447}

\bibitem[\protect\citeauthoryear{{Masci} et~al.,}{{Masci}
  et~al.}{2019}]{ztf2019}
{Masci} F.~J.,  et~al., 2019, \mn@doi [\pasp] {10.1088/1538-3873/aae8ac}, \href
  {https://ui.adsabs.harvard.edu/abs/2019PASP..131a8003M} {131, 018003}

\bibitem[\protect\citeauthoryear{{Mata S{\'a}nchez} et~al.,}{{Mata S{\'a}nchez}
  et~al.}{2023}]{matasanchez2023}
{Mata S{\'a}nchez} D.,  et~al., 2023, \mn@doi [\mnras] {10.1093/mnras/stad203},
  \href {https://ui.adsabs.harvard.edu/abs/2023MNRAS.520.2217M} {520, 2217}

\bibitem[\protect\citeauthoryear{{Medvedev}, {Gilfanov}, {Sazonov}, {Sunyaev}
  \& {Khorunzhev}}{{Medvedev} et~al.}{2022}]{eROSITA-var-sources-2022}
{Medvedev} P.~S.,  {Gilfanov} M.~R.,  {Sazonov} S.~Y.,  {Sunyaev} R.~A.,
  {Khorunzhev} G.~A.,  2022, \mn@doi [Astronomy Letters]
  {10.1134/S1063773722120015}, \href
  {https://ui.adsabs.harvard.edu/abs/2022AstL...48..735M} {48, 735}

\bibitem[\protect\citeauthoryear{{Miller} et~al.,}{{Miller}
  et~al.}{2020}]{miller2020}
{Miller} J.~M.,  et~al., 2020, \mn@doi [\apj] {10.3847/1538-4357/abbb2e}, \href
  {https://ui.adsabs.harvard.edu/abs/2020ApJ...904...49M} {904, 49}

\bibitem[\protect\citeauthoryear{{Papitto} et~al.,}{{Papitto}
  et~al.}{2013}]{papitto2013}
{Papitto} A.,  et~al., 2013, \mn@doi [\nat] {10.1038/nature12470}, \href
  {https://ui.adsabs.harvard.edu/abs/2013Natur.501..517P} {501, 517}

\bibitem[\protect\citeauthoryear{{Predehl} et~al.,}{{Predehl}
  et~al.}{2021}]{erosita2021}
{Predehl} P.,  et~al., 2021, \mn@doi [\aap] {10.1051/0004-6361/202039313},
  \href {https://ui.adsabs.harvard.edu/abs/2021A&A...647A...1P} {647, A1}

\bibitem[\protect\citeauthoryear{{Ray} et~al.,}{{Ray} et~al.}{2020}]{ray2020}
{Ray} P.~S.,  et~al., 2020, \mn@doi [Research Notes of the American
  Astronomical Society] {10.3847/2515-5172/ab7eb5}, \href
  {https://ui.adsabs.harvard.edu/abs/2020RNAAS...4...37R} {4, 37}

\bibitem[\protect\citeauthoryear{{Roberts}}{{Roberts}}{2013}]{roberts2013}
{Roberts} M. S.~E.,  2013, in {van Leeuwen} J.,  ed., ~ Vol. 291, Neutron Stars
  and Pulsars: Challenges and Opportunities after 80 years. pp 127--132
  (\mn@eprint {arXiv} {1210.6903}), \mn@doi{10.1017/S174392131202337X}

\bibitem[\protect\citeauthoryear{{Romani}}{{Romani}}{2012}]{romain2012}
{Romani} R.~W.,  2012, \mn@doi [\apjl] {10.1088/2041-8205/754/2/L25}, \href
  {https://ui.adsabs.harvard.edu/abs/2012ApJ...754L..25R} {754, L25}

\bibitem[\protect\citeauthoryear{{Romani}, {Kandel}, {Filippenko}, {Brink}  \&
  {Zheng}}{{Romani} et~al.}{2021}]{romani2021}
{Romani} R.~W.,  {Kandel} D.,  {Filippenko} A.~V.,  {Brink} T.~G.,   {Zheng}
  W.,  2021, \mn@doi [\apjl] {10.3847/2041-8213/abe2b4}, \href
  {https://ui.adsabs.harvard.edu/abs/2021ApJ...908L..46R} {908, L46}

\bibitem[\protect\citeauthoryear{{Romani}, {Kandel}, {Filippenko}, {Brink}  \&
  {Zheng}}{{Romani} et~al.}{2022}]{romani2022}
{Romani} R.~W.,  {Kandel} D.,  {Filippenko} A.~V.,  {Brink} T.~G.,   {Zheng}
  W.,  2022, \mn@doi [\apjl] {10.3847/2041-8213/ac8007}, \href
  {https://ui.adsabs.harvard.edu/abs/2022ApJ...934L..17R} {934, L17}

\bibitem[\protect\citeauthoryear{{Roy} et~al.,}{{Roy} et~al.}{2015}]{roy2015}
{Roy} J.,  et~al., 2015, \mn@doi [\apjl] {10.1088/2041-8205/800/1/L12}, \href
  {https://ui.adsabs.harvard.edu/abs/2015ApJ...800L..12R} {800, L12}

\bibitem[\protect\citeauthoryear{{Salvetti} et~al.,}{{Salvetti}
  et~al.}{2017}]{salvetti2017}
{Salvetti} D.,  et~al., 2017, \mn@doi [\mnras] {10.1093/mnras/stx1247}, \href
  {https://ui.adsabs.harvard.edu/abs/2017MNRAS.470..466S} {470, 466}

\bibitem[\protect\citeauthoryear{{Scargle}}{{Scargle}}{1982}]{scargle1982}
{Scargle} J.~D.,  1982, \mn@doi [\apj] {10.1086/160554}, \href
  {https://ui.adsabs.harvard.edu/abs/1982ApJ...263..835S} {263, 835}

\bibitem[\protect\citeauthoryear{{Schmidt} et~al.,}{{Schmidt}
  et~al.}{2019}]{Schmidt2019}
{Schmidt} S.~J.,  et~al., 2019, \mn@doi [\apj] {10.3847/1538-4357/ab148d},
  \href {https://ui.adsabs.harvard.edu/abs/2019ApJ...876..115S} {876, 115}

\bibitem[\protect\citeauthoryear{{Smith} et~al.,}{{Smith}
  et~al.}{2023}]{FermiPSRs2023}
{Smith} D.~A.,  et~al., 2023, arXiv e-prints, \href
  {https://ui.adsabs.harvard.edu/abs/2023arXiv230711132S} {p. arXiv:2307.11132}

\bibitem[\protect\citeauthoryear{{Strader} et~al.,}{{Strader}
  et~al.}{2019}]{strader2019}
{Strader} J.,  et~al., 2019, \mn@doi [\apj] {10.3847/1538-4357/aafbaa}, \href
  {https://ui.adsabs.harvard.edu/abs/2019ApJ...872...42S} {872, 42}

\bibitem[\protect\citeauthoryear{{Sunyaev} et~al.,}{{Sunyaev}
  et~al.}{2021}]{Sunyaev2021}
{Sunyaev} R.,  et~al., 2021, \mn@doi [\aap] {10.1051/0004-6361/202141179},
  \href {https://ui.adsabs.harvard.edu/abs/2021A&A...656A.132S} {656, A132}

\bibitem[\protect\citeauthoryear{{Swihart} et~al.,}{{Swihart}
  et~al.}{2020}]{swihart2020}
{Swihart} S.~J.,  et~al., 2020, \mn@doi [\apj] {10.3847/1538-4357/ab77ba},
  \href {https://ui.adsabs.harvard.edu/abs/2020ApJ...892...21S} {892, 21}

\bibitem[\protect\citeauthoryear{{Swihart}, {Strader}, {Aydi}, {Chomiuk},
  {Dage}  \& {Shishkovsky}}{{Swihart} et~al.}{2021}]{swihart2021}
{Swihart} S.~J.,  {Strader} J.,  {Aydi} E.,  {Chomiuk} L.,  {Dage} K.~C.,
  {Shishkovsky} L.,  2021, \mn@doi [\apj] {10.3847/1538-4357/abe1be}, \href
  {https://ui.adsabs.harvard.edu/abs/2021ApJ...909..185S} {909, 185}

\bibitem[\protect\citeauthoryear{{Swihart}, {Strader}, {Chomiuk}, {Aydi},
  {Sokolovsky}, {Ray}  \& {Kerr}}{{Swihart} et~al.}{2022}]{swihart2022}
{Swihart} S.~J.,  {Strader} J.,  {Chomiuk} L.,  {Aydi} E.,  {Sokolovsky} K.~V.,
   {Ray} P.~S.,   {Kerr} M.,  2022, \mn@doi [\apj] {10.3847/1538-4357/aca2ac},
  \href {https://ui.adsabs.harvard.edu/abs/2022ApJ...941..199S} {941, 199}

\bibitem[\protect\citeauthoryear{{Turchetta}, {Linares}, {Koljonen}  \&
  {Sen}}{{Turchetta} et~al.}{2023}]{PSR1622arXiv}
{Turchetta} M.,  {Linares} M.,  {Koljonen} K.,   {Sen} B.,  2023, \mn@doi
  [arXiv e-prints] {10.48550/arXiv.2307.06330}, \href
  {https://ui.adsabs.harvard.edu/abs/2023arXiv230706330T} {p. arXiv:2307.06330}

\bibitem[\protect\citeauthoryear{{Warner}}{{Warner}}{1995}]{warner1995}
{Warner} B.,  1995, {Cataclysmic variable stars}.
~ Vol. 28, Cambridge University Press

\bibitem[\protect\citeauthoryear{{Warner} \& {Woudt}}{{Warner} \&
  {Woudt}}{2009}]{2009MNRAS.397..979W}
{Warner} B.,  {Woudt} P.~A.,  2009, \mn@doi [\mnras]
  {10.1111/j.1365-2966.2009.15006.x}, \href
  {https://ui.adsabs.harvard.edu/abs/2009MNRAS.397..979W} {397, 979}

\bibitem[\protect\citeauthoryear{{Wilms}, {Allen}  \& {McCray}}{{Wilms}
  et~al.}{2000}]{wilms2000}
{Wilms} J.,  {Allen} A.,   {McCray} R.,  2000, \mn@doi [\apj] {10.1086/317016},
  \href {https://ui.adsabs.harvard.edu/abs/2000ApJ...542..914W} {542, 914}

\bibitem[\protect\citeauthoryear{{Yap}, {Kong}  \& {Li}}{{Yap}
  et~al.}{2023}]{yap2023}
{Yap} Y.~X.~J.,  {Kong} A.,   {Li} K.~L.,  2023, \mn@doi [arXiv e-prints]
  {10.48550/arXiv.2307.13482}, \href
  {https://ui.adsabs.harvard.edu/abs/2023arXiv230713482Y} {p. arXiv:2307.13482}

\bibitem[\protect\citeauthoryear{{Zasche}, {Wolf}  \& {Vra{\v{s}}til}}{{Zasche}
  et~al.}{2017}]{zasche2017}
{Zasche} P.,  {Wolf} M.,   {Vra{\v{s}}til} J.,  2017, \mn@doi [\mnras]
  {10.1093/mnras/stx989}, \href
  {https://ui.adsabs.harvard.edu/abs/2017MNRAS.469.2952Z} {469, 2952}

\bibitem[\protect\citeauthoryear{{Zharikov}, {Tovmassian}, {Aviles}, {Michel},
  {Gonzalez-Buitrago}  \& {Garc{\'\i}a-D{\'\i}az}}{{Zharikov}
  et~al.}{2013}]{zharikov2013}
{Zharikov} S.,  {Tovmassian} G.,  {Aviles} A.,  {Michel} R.,
  {Gonzalez-Buitrago} D.,   {Garc{\'\i}a-D{\'\i}az} M.~T.,  2013, \mn@doi
  [\aap] {10.1051/0004-6361/201220099}, \href
  {https://ui.adsabs.harvard.edu/abs/2013A&A...549A..77Z} {549, A77}

\bibitem[\protect\citeauthoryear{{Zharikov}, {Kirichenko}, {Zyuzin}, {Shibanov}
   \& {Deneva}}{{Zharikov} et~al.}{2019}]{zharikov2019}
{Zharikov} S.,  {Kirichenko} A.,  {Zyuzin} D.,  {Shibanov} Y.,   {Deneva}
  J.~S.,  2019, \mn@doi [\mnras] {10.1093/mnras/stz2475}, \href
  {https://ui.adsabs.harvard.edu/abs/2019MNRAS.489.5547Z} {489, 5547}

\makeatother
\end{thebibliography}







\bsp	
\label{lastpage}
\end{document}